\RequirePackage{rotating}
\documentclass[a4paper,fleqn,usenatbib]{mnras}
\usepackage{newtxtext,newtxmath}
\usepackage[T1]{fontenc}
\usepackage{ae,aecompl}

\usepackage{color}
\usepackage{amsmath}
\usepackage{amssymb}
\usepackage{graphicx}
\usepackage{epsfig, subfigure}
\usepackage{booktabs}
\usepackage{array}
\usepackage{rotating}
\usepackage{caption} 
\usepackage{tabularx}
\usepackage{bm}

\newcommand{\ra}[1]{\renewcommand{\arraystretch}{#1}}

\newcommand{\WFIRST}{{\slshape WFIRST}}
\newcommand{\buzzard}{{\textsc{Buzzard-v1.1}}}
\newcolumntype{P}[1]{>{\centering\arraybackslash}p{#1}}

\newcommand{\eg}{e.g.}
\newcommand{\ie}{i.e.}

\hypersetup{draft}

\title[WFIRST Line Blending]{Effects of [N{\,\sc ii}] and H$\alpha$ Line Blending on the \WFIRST\ Galaxy Redshift Survey}

\author[Martens et al.]{
Daniel Martens,$^{1}$
Xiao Fang,$^{1,2}$
M. A. Troxel,$^{1}$
Joe DeRose,$^{3,4}$,\newauthor
Christopher M. Hirata,$^{1}$ 
Risa H. Wechsler,$^{3,4,5}$
and Yun Wang$^{6}$
\\
$^{1}$Center for Cosmology and AstroParticle Physics, Department of Physics, The Ohio State University, 191 W Woodruff Ave, \\  Columbus OH 43210, U.S.A.\\
$^{2}$Department of Astronomy and Steward Observatory, University of Arizona, 933 N Cherry Ave, Tucson, AZ 85719, U.S.A.\\
$^{3}$Department of Physics, Stanford University, 382 Via Pueblo Mall, Stanford, CA 94305, USA\\
$^{4}$Kavli Institute for Particle Astrophysics \& Cosmology, P. O. Box 2450, Stanford University, Stanford, CA 94305, USA\\
$^{5}$SLAC National Accelerator Laboratory, Menlo Park, CA 94025, USA\\
$^{6}$IPAC, Mail Code 314-6, California Institute of Technology, 1200 East California Boulevard, Pasadena, CA 91125, USA
}

\date{Accepted XXX. Received YYY; in original form ZZZ}

\pubyear{2017}

\begin{document}
\label{firstpage}
\pagerange{\pageref{firstpage}--\pageref{lastpage}}
\maketitle

\begin{abstract}
The Wide Field Infrared Survey Telescope (\WFIRST) will conduct a galaxy redshift survey using the H$\alpha$ emission line primarily for spectroscopic redshift determination.
Due to the modest spectroscopic resolution of the grism, the H$\alpha$ and the neighboring [N{\,\sc ii}] lines are blended, leading to a redshift bias that depends on the [N{\,\sc ii}]/H$\alpha$ ratio, which is correlated with a galaxy's metallicity, hence mass and ultimately environment.
We investigate how this bias propagates into the galaxy clustering and cosmological parameters obtained from the \WFIRST. Using simulation, we explore the effect of line blending on redshift-space distortion and baryon acoustic oscillation (BAO) measurements. We measure the BAO parameters $\alpha_{\parallel}$, $\alpha_{\perp}$, the logarithmic growth factor $f_{v}$, and calculate their errors based on the correlations between the line ratio and large-scale structure. We find $\Delta\alpha_{\parallel} = 0.31 \pm 0.23 \%$ ($0.26\pm0.17\%$), $\Delta\alpha_{\perp} = -0.10\pm0.10\%$ ($-0.12 \pm 0.11 \%$), and $\Delta f_{v} = 0.17\pm0.33\%$ ($-0.20 \pm 0.30\%$) for redshift 1.355--1.994 (0.700--1.345), which use approximately 18$\%$, 9$\%$, and 7$\%$ of the systematic error budget in a root-sum-square sense. These errors may already be tolerable but further mitigations are discussed.
Biases due to the environment-independent redshift error can be mitigated by measuring the redshift error probability distribution function. High-spectral-resolution re-observation of a few thousand galaxies would be required (if by direct approach) to reduce them to below 25$\%$ of the error budget.
Finally, we outline the next steps to improve the modeling of [N{\,\sc ii}]-induced blending biases and their interaction with other redshift error sources.
\end{abstract}



\section{Introduction}
\label{sec:intro}

Observations of high-redshift supernovae in the 1990s provided the first direct evidence for an acceleration in the expansion rate of the universe \citep{riess1998,perlmutter1999}. Whatever field or particle is responsible for this surprising acceleration has been dubbed ``dark energy,'' and one of the major observational programs in modern cosmology is to measure its properties. It is of particular interest to determine whether the dark energy is consistent with a cosmological constant; whether it requires new dynamical degrees of freedom; or whether cosmic acceleration arises instead from a modification to the laws of gravity on large scales. 

In order to measure the properties of dark energy, cosmologists employ a variety of techniques. Observations of supernovae provide information on the expansion rate of the Universe for different redshifts, probing the effects of dark energy throughout cosmic history. Weak lensing surveys probe the matter distribution, allowing measurements of clustering at various redshifts. Galaxy clusters are the most massive collapsed objects produced by cosmological structure formation, and can be traced using a wide range of observables (the visible galaxy content, the hot gas via X-ray emission or the Sunyaev-Zel'dovich effect, and weak lensing). Galaxy redshift surveys -- the subject of this paper -- can trace cosmic structures in three dimensions, although their cosmological interpretation requires accurate modeling of the relation between the visible galaxies and the mostly unseen matter.

The size of redshift surveys have been steadily increasing in tandem with technological improvements. 
A sample of over 200,000 galaxies was investigated in the 2dF Galaxy Redshift Survey (2dFGRS), constraining cosmological parameters within specific cosmological models \citep{cole2005}. This was followed by the 6dF Galaxy Survey \citep{jones2009}. A spectroscopic analysis of over 54,000 luminous red galaxies using the Sloan Digital Sky Survey (SDSS) found evidence for baryon acoustic oscillations, and provided additional constraints on the cosmological parameters \citep{eisenstein2005}. Further analysis has been done combining SDSS with 2dFGRS, as well as analyzing data provided by SDSS DR7 \citep{percival2007,percival2010}. More recently, the WiggleZ Dark Energy Survey has been used to measure the BAO peak at different redshifts \citep{blake2011}, while others have analyzed the distance to these redshifts \citep{xu2012}. The SDSS-III/BOSS project, which included an upgraded spectrograph with enhanced red sensitivity, collected spectra of over 2.4 million galaxies \citep{alam2015}. The redshift range of SDSS-III is extended by SDSS-IV (eBOSS) which is currently observing \citep{dawson2016,blanton2017}.

This progress is expected to continue: the Taipan survey will look at low-redshift galaxies, over half the sky \citep{cunha2017}, and the Dark Energy Spectroscopic Instrument (DESI) will conduct a comprehensive spectroscopic survey of galaxies and quasars over the Northern sky \citep{DESI1,DESI2}. A substantially deeper survey is planned by the Prime Focus Spectrograph (PFS) \citep{tamura2016}, which will extend ground-based spectroscopic coverage out to 1.26 $\mu$m, and the 4m Multi Object Spectroscopic Telescope \citep[4MOST][]{depagne2015} project will conduct optical spectroscopy in the south. The {\slshape Euclid} mission will conduct a space-based near-infrared grism survey with portions of its footprint in both hemispheres \citep{laureijs2011}.


The Wide Field InfraRed Survey Telescope (\WFIRST) will be a 2.4 m space telescope that carries out a wide range of investigations in cosmology, exoplanets, and other areas of astrophysics \citep{dressler2012,green2012,spergel2015} following its launch in the mid-2020's. \WFIRST\ will carry a Wide Field Instrument (WFI; capable of imaging and slitless spectroscopy) and a coronagraph. The galaxy redshift survey program on \WFIRST\ will use the grism to observe emission lines in the 1.00--1.93 $\mu$m bandpass\footnote{This wavelength range was chosen for the System Requirements Review/Mission Definition Review, and is somewhat different from that considered during previous iterations of the \WFIRST\ design.} and obtain redshifts for $1.8\times 10^6$ galaxies per month of observations. The principal tracer of large scale structure will be the H$\alpha$ emission line (at 6565 \AA), which is visible in \WFIRST\ out to a maximum redshift of $z=1.94$; at higher redshift, other emission lines will be used, most notably the [O{\,\sc iii}] doublet (4960 and 5008 \AA), which is visible out to $z=2.85$.

The grism spectroscopy technique has the advantage of simplicity (the grism occupies one slot on the \WFIRST\ filter wheel, with no additional moving parts); it provides enormous multiplexing; and it does not require that targets be selected in advance (thus providing operational flexibility, and avoiding selection biases that are ``baked in'' to traditional redshift surveys at the time of target selection). However, it does have drawbacks. One is that without a slit, each pixel is exposed to the full dispersed sky scene rather than only the targeted galaxy, which leads to higher backgrounds and confusion from other sources. Grism spectroscopy also has some constraints: \WFIRST\ requires a wide-field grism in a converging beam; it was a significant design challenge for all field positions and all wavelengths to focus simultaneously, and solutions are only available at moderate spectral resolution \citep{2016SPIE.9904E..12G}. At this resolution ($R\sim 690$ per 2-pixel element at $\lambda = 1.5$ $\mu$m), the H$\alpha$ line is partially blended with the neighboring [N{\,\sc ii}] lines. This is a similar situation to the WFC3 Infrared Spectroscopic Parallel (WISP) Survey, which used an even lower-resolution grism on the {\slshape Hubble Space Telescope} where H$\alpha$+[N{\,\sc ii}] was completely blended \citep{atek2010}. The results of the WISP survey therefore have been used to make predictions on the effectiveness of future surveys \citep{colbert2013,2015ApJ...811..141M,2018MNRAS.474..177M}, particularly through the discussion of the changes in the luminosity function of galaxies when line blending issues are present. In the past, a correction factor of 0.71 has been applied to these H$\alpha$ luminosities of galaxies to account for [N{\,\sc ii}] contamination \citep{mehta2015}\footnote{Other studies have used different methods to account for the [N{\,\sc ii}] contamination. For example, \cite{2013MNRAS.428.1128S} uses a relation between [N{\,\sc ii}]/H$\alpha$ ratio and [N{\,\sc ii}]+H$\alpha$ line equivalent width as described in \cite{2008ApJ...677..169V} to de-blend their [N{\,\sc ii}]+H$\alpha$ fluxes derived from narrow-band observations.}. Further analyses of the effect of blending on the luminosity function, as well as an empirical parameterization of the [N{\,\sc ii}]/H$\alpha$ flux ratio as a function of galaxy properties and redshift, were completed by \cite{faisst2017}.

A separate concern arising from the line blending effect is the change in the observed redshift. Since the [N{\,\sc ii}] lines are asymmetric in emission strength, they will shift the ``observed'' redshift $z_{\rm obs}$ (assuming that the line centroid is at the H$\alpha$ wavelength in the rest frame), to be redder than the true redshift, $z_{\rm true}$ \citep{faisst2017}. Since the line ratio is determined by H{\,\sc ii} region physics and depends on metallicity, this effect is not the same for all galaxies: it is not removable by subtracting a mean bias. Furthermore, since metallicity is correlated with galaxy mass and hence with large-scale structure, H$\alpha$+[N{\,\sc ii}] line blending leads to a redshift bias that is correlated with large-scale structure and could have highly non-trivial effects on the inferred cosmological parameters.

The purpose of this paper is to quantify the extent of this observational problem, understand how it will affect the measured galaxy correlation function if left unmitigated, and discuss potential mitigation strategies, to the extent that they will be necessary. We will analyze a sample of $>10^8$ galaxies using mock data from the \buzzard\ simulation. We compute the correlation function and perform BAO and RSD fits with both the true redshift catalog and the observed redshift catalog and assess the differences. We also create a ``shuffled'' redshift catalog, where the redshift errors $\Delta z/(1+z)$ are scrambled, which allows us to test which changes in the clustering properties are due to the correlations of redshift error with large-scale structure, and which are due to the distribution of redshift errors alone.

The outline of the paper is as follows. In Section \ref{sec:objectives}, we discuss the line blending problem in more detail, and introduce the main questions which this paper will attempt to answer. In Section \ref{sec:outline}, we describe the general road map to answering these questions, discussing the analysis strategies we will use to dissect the simulation results. We discuss details of our simulation in Section \ref{sec:simulation}, as well as the description of the catalogs. In Section \ref{sec:fitting}, we discuss our fitting strategies and methods, and display our fits for redshift space distortion and BAO parameters to the data. In Section \ref{sec:mitigation} we discuss our results and how they compare to the requirements for \WFIRST, with a brief discussion of possible mitigation strategies. We conclude in section \ref{sec:conclusion}.

For this analysis, we use a flat $\Lambda$CDM cosmology with parameters of $\Omega_{m} = 0.286$, $\Omega_{\Lambda} = 0.714$, $\sigma_{8} = 0.82$, $h = 0.7$, $\Omega_{b} = 0.047$, and $n_{s} = 0.96$. This is consistent with the parameters used in the generation of the \buzzard\ simulations. Finally, all line wavelengths are referenced to their vacuum values.

\section{Objectives}
\label{sec:objectives}

All spectroscopic galaxy surveys contain some redshift errors, in the sense that the observed redshift $z_{\rm obs}$ deviates from the true redshift $z_{\rm true}$. Among existing samples used for large-scale structure analyses, this is most evident for quasars \citep[e.g.][]{dawson2016}, since the most accessible lines are broad and often asymmetric, and redshift errors of several hundred km s$^{-1}$ are common.

In the case of \WFIRST, photon noise in the centroid of the emission lines will be the dominant source of scatter in the $z_{\rm obs}$ vs.\ $z_{\rm true}$ relation. The inclusion of statistical noise results in both a decrease in the constraining power of the survey, and the suppression of power at large $k_\parallel$. This noise does not require any additional treatment to remove, since the central values are unchanged. However, in the \WFIRST\ survey, blending of the H$\alpha$ emission line with the neighboring [N{\,\sc ii}] doublet (one doublet member is on each side of H$\alpha$) results in an offset $z_{\rm obs} > z_{\rm true}$ for objects of higher [N{\,\sc ii}]/H$\alpha$ ratio (which are likely higher metallicity and hence probably found in denser environments). Other possible errors could involve the emission line strength and the angular size of a galaxy affecting the width of the $z_{\rm obs}-z_{\rm true}$ distribution, and this may in turn affect the redshift offset, if it interacts with point spread function (PSF) asymmetry. Both of these properties may be correlated with the galaxy environment. They are, however, beyond the scope of this paper; we plan to revisit PSF asymmetry and other instrument-related issues when the \WFIRST\ grism simulation pipeline is at a more advanced stage.

\begin{figure}
\centering
\includegraphics[width=\columnwidth,height=!,keepaspectratio]{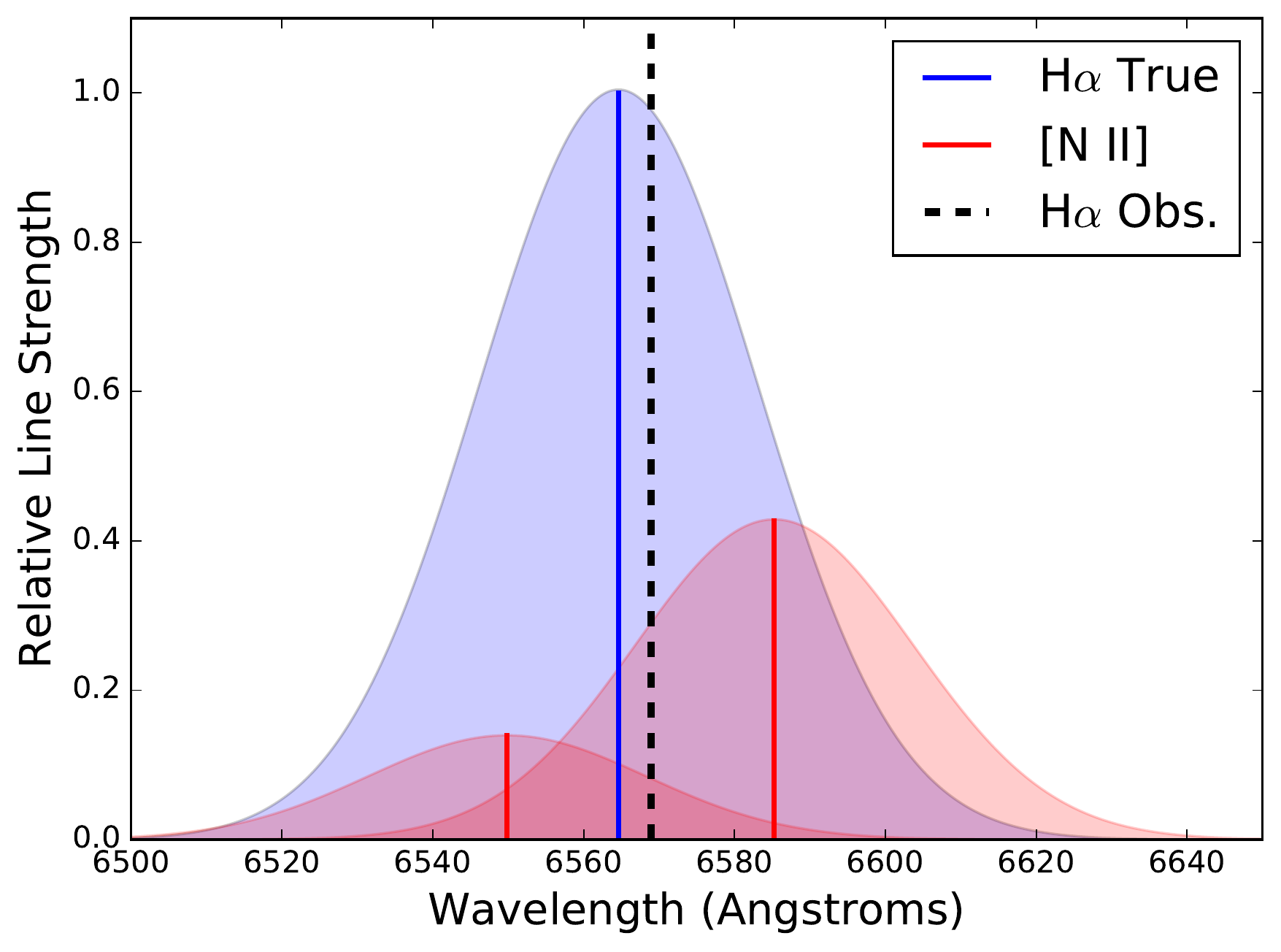}
\caption{A visual representation of the [N{\,\sc ii}] lines and H$\alpha$ line for the line blending scenario. The difference in strength between the two nitrogen lines is constant for each galaxy with a ratio of approximately 0.32, while the difference between the larger [N{\,\sc ii}] line and the H$\alpha$ line varies depending on the metallicity of the galaxy in question. Here, we show the median difference of our sample, with a ratio of [N{\,\sc ii}]6585/H$\alpha$ of 0.427. The Gaussian spreading centered at each line has a standard deviation of $\sigma_{\mathrm{grism}}$, calculated generally in Eq. \ref{eq:grism_spread}. In this example, we have shown the spreading due to a galaxy with radius 4 kpc, at redshift 1.5. The black dotted line at 6569\r{A} (about 4.3\r{A} larger than the true H$\alpha$ wavelength, corresponding to an increase in redshift of about $6.6\times 10^{-4}$) is the resulting observed line, given that the constituent lines cannot be resolved. 
}
\label{fig:blending}
\end{figure}

Figure~\ref{fig:blending} demonstrates the scale of the line blending problem. The \WFIRST\ grism has a spectral resolving power of
\begin{equation} \label{eq:grism_spread}
\frac{\lambda}{\Delta \lambda} = 461 \left( \frac{\lambda_{\rm obs}}{1\,\mu\rm m} \right),
\end{equation}
where $\lambda$ is the observed wavelength (this is measured for a 2-pixel element; an extended galaxy will be bigger and hence have lower spectral resolution).\footnote{See \citet{gehrels2014}; updated information can be found at the \WFIRST\ project website:\\ \tt https://wfirst.gsfc.nasa.gov/science/WFIRST\_Reference\_ Information.html} It is seen that at this resolution, the H$\alpha$ line and surrounding [N{\,\sc ii}] lines are blended, and a fit to a single line will find something close to the centroid of the blended features. If there was a known error probability distribution function (PDF) $P(z_{\rm obs}-z_{\rm true}|z_{\rm true})$ that was both uncorrelated with galaxy environment and valid for every type of galaxy in the sample, then we could incorporate this in the theoretical correlation function, and the only effect of the redshift errors due to line blending would be a reduction in the statistical constraining power in the survey. If, however, the redshift error PDF is either not known or is correlated with galaxy environment, further steps may be needed to maximize accurate redshift reconstruction. Such correlations may be problematic for a survey even if the systematic redshift error is small compared to the statistical errors for one single galaxy.

In this paper, we use the \buzzard\ simulation (described in Sec. \ref{subsec:catalog_gen}) to address several key questions regarding the impact of the line blending phenomenon on the \WFIRST\ redshift survey: 
\begin{enumerate}
\item If we ignore the effects of line blending, what biases are induced in the baryon acoustic oscillation (BAO) and redshift space distortion (RSD) parameters? How does this compare to the corresponding statistical errors on these parameters, or the errors required by the \WFIRST\ Science Requirements Document (SRD)?
\item Do we need to mitigate biases caused by correlations between galaxy environment and redshift offsets due to line blending?
\item If we determine that the problems are significant enough to require some level of mitigation, what type of mitigations are necessary?
\end{enumerate}


The \WFIRST\ SRD defines required performance based on a Reference Survey of 0.70 years\footnote{The actual allocations will be determined in the future by the Implementation SWG.}, with $1\sigma$ statistical errors of 0.70\%\ on the transverse BAO distance scale ($\alpha_\perp$); 1.28\%\ on the radial BAO distance scale ($\alpha_\parallel$); and 1.28\%\ on the rate of growth of structure measured from RSD ($f_v$). Observational systematic errors are allocated an error of 0.58 times the Reference Survey statistical errors, i.e.\ 0.41\%\ ($\alpha_\perp$), 0.74\%\ ($\alpha_\parallel$), and 0.74\%\ ($f_v$).\footnote{This means that the combination of statistical errors and observational systematic errors would be $\sqrt{1+0.58^2}=1.16$ times larger than the statistical errors alone. Note furthermore that the SRD allows for other sources of error as well.} Note that these are {\em requirements} -- it is always desirable to have smaller systematic errors, but if they exceed their allocation they must be mitigated. Finally, note that the H$\alpha$+[N{\,\sc ii}] blending is only one type of observational systematic error, and as such should consume only a fraction of the systematic error budget (the exact percentage has not been set; part of the purpose of this paper is to inform this discussion). Other systematic errors include uncertainties in the optical distortion models (both for grism mode and the direct imaging used as a reference); wavelength calibration; detector effects (e.g., flat fielding, cross-talk); and point spread function asymmetry (e.g., centroid fitting in the presence of coma). The draft error budget for these in the SRD is 0.23 times the reference statistical errors.

\section{Calculation Outline}
\label{sec:outline}

\begin{figure*}
\centering
\includegraphics[scale=0.45]{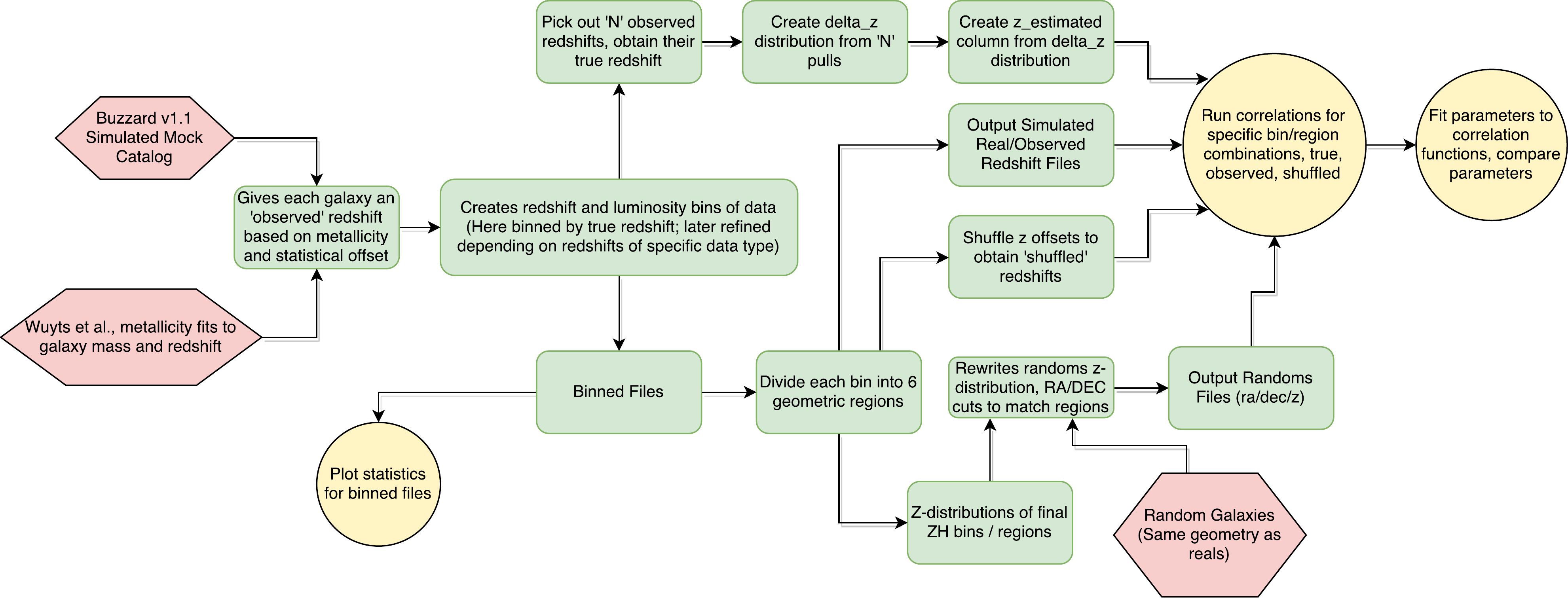}
\caption{A flow chart describing the simulation pipeline; see Sec. \ref{sec:outline} for details. The red hexagons indicate steps where data is input into the process, and yellow circles indicate steps where we output statistics, correlation functions, or fit parameters. Also included in the flowchart is a method for possible analysis of the ability of a ground-based spectroscopic observation to reproduce the observed catalog; although it is not implemented in this work, we discuss its potential in Sec. \ref{sec:mitigation}.}
\label{fig:flowchart}
\end{figure*}

In this section, we explain the specific steps taken throughout the analysis in order to fully understand the detailed effects of line blending. Figure \ref{fig:flowchart} shows a flowchart of the process, beginning with the \buzzard\ mock galaxy catalog, moving through the calculations we perform, and ending with our parameter estimates. Although more technical details for each process will be expanded later on in Section \ref{sec:simulation}, we take a moment here to give a high-level overview of the entire pipeline.

We begin with the mock catalog from the \buzzard\ simulation. We use a ``true'' redshift for each galaxy in the catalog that incorporates peculiar velocity effects, but does not yet contain any line blending or  statistical errors. The catalog and its generation are described in detail in Section \ref{subsec:catalog_gen}. The catalog covers one quadrant of the sky ($\pi$ steradians).

Our first step is to assign to each galaxy a [N{\,\sc ii}]/H$\alpha$ ratio based upon that galaxy's redshift and stellar mass. Once each galaxy has a [N{\,\sc ii}]/H$\alpha$ ratio, we can then calculate the observed redshift $z_{\mathrm{obs}}$ for each galaxy, which incorporates the effect of line blending. We also include a statistical offset of the redshift due to photon noise in the center of the line. These redshifts form the ``observed redshift'' catalog.

We next create a separate redshift for each galaxy called the ``shuffled'' redshift, or $z_{\mathrm{shf}}$. The purpose of this is to provide a redshift catalog where the distribution of the observed redshift catalog is accounted for, but removes any correlation between the [N{\,\sc ii}]/H$\alpha$ ratio and the galaxy environment. By later comparing clustering parameters from the ``shuffled'' redshifts to those from the observed redshifts, we can determine how much of the effect we see is captured in the 1-point distribution of redshift errors, and how much depends on environmental correlations.

\begin{figure}
\centering
\includegraphics[width=\columnwidth,height=!,keepaspectratio]{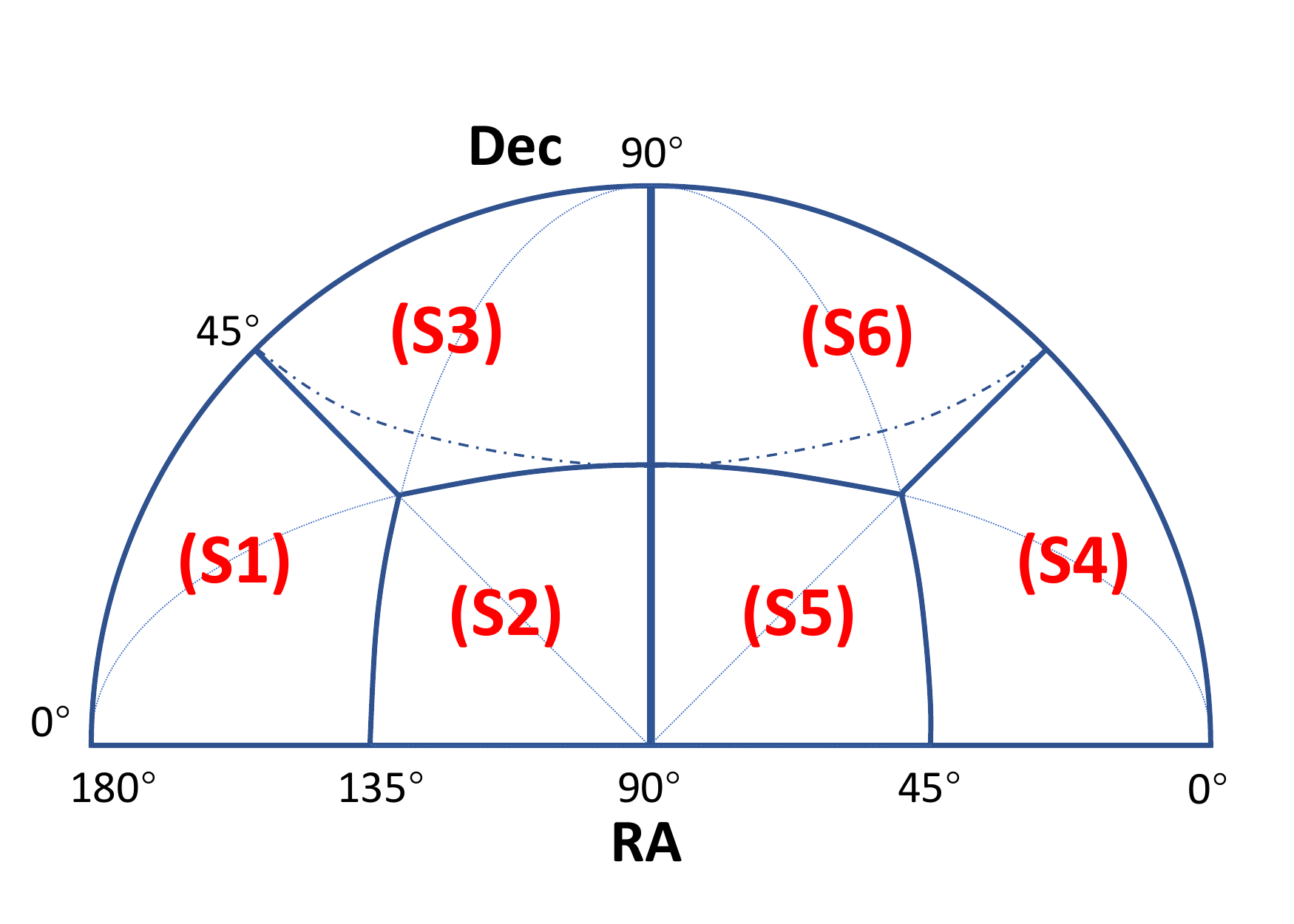}
\caption{We display the 6 separate regions on the sky, within which we have independently calculated the correlation function, as discussed in Sec. \ref{subsec:binning}. Note that although we must display a 2D projection, the sectors are congruent on the sphere. (The simulation does not cover the actual \WFIRST\ footprint, which is likely to be placed in the Southern Hemisphere, but this does not matter in a statistically isotropic universe.)}
\label{fig:regions}
\end{figure}

Next, the galaxies are binned, depending on the galaxy redshift (Z2, Z3) and line flux (H1, H2, H3).  The group cuts are described in Table \ref{tab:binning2}. Note that a given galaxy may be in one bin in the true redshift catalog, but in another bin in the observed redshift catalog, if the offsets push the observed redshift into another bin. Furthermore, to ease the computational burden, each bin is split into six congruent kite-shaped geometric regions (S1, S2, ... S6) on the sky, each of solid angle $\pi/6$ steradians, as shown in Figure \ref{fig:regions}. These sections are counted separately, and then recombined in the analysis of the correlation functions. 

For each redshift/flux bin (which will be referred to as ``ZH'' bins) and geometric region, we fit redshift-space distortion (RSD) parameters and baryon acoustic oscillation (BAO) parameters, and compare the resulting parameter shifts with the \WFIRST\ error budget to assess their significance and the possible need for mitigation.

\section{Simulation}
\label{sec:simulation}

\subsection{Catalog generation}
\label{subsec:catalog_gen}

We make use of the \buzzard\ mock galaxy catalog that we describe briefly here and refer the interested reader to more detailed descriptions in \citet{DeRose2019} and \citet{Wechsler2019}. \buzzard\ is a simulated galaxy catalog constructed from a set of three nested dark matter-only lightcone simulations which are progressively lower resolution at higher redshifts. The lightcones have volumes of $(1050 ~h^{-1}\textrm{Mpc} )^3$, $(2600 ~h^{-1}\textrm{Mpc})^3$, $(4000 ~h^{-1}\textrm{Mpc} )^3$, particle masses of $2.7\times10^{10}~h^{-1}\textrm{M}_\odot $, $1.3\times10^{11}~h^{-1}\textrm{M}_\odot$, $4.8\times10^{11}~h^{-1}\textrm{M}_\odot$ and force softenings of $20 ~h^{-1} \textrm{kpc}$, $35 ~h^{-1}\textrm{kpc}$, $53 ~h^{-1}\textrm{kpc}$ respectively. The highest resolution $(1050 ~h^{-1}\textrm{Mpc} )^3$ simulation is used for $z<0.34$, the $(2600 ~h^{-1}\textrm{Mpc})^3$ for $0.34<=z<0.9$ and the $(4000 ~h^{-1}\textrm{Mpc})^3$ simulation for $0.9<z<2.35$. These simulations are run using \textsc{L-GADGET2}, a version of \textsc{GADGET2} \citep{Springel2005} modified for memory efficiency with 2nd-order Lagrangian perturbation theory (2LPT) initial conditions created using \textsc{2LPTIC} \citep{Crocce2006}. Lightcones are generated on the fly as the simulations run.

\begin{figure}
\centering
\includegraphics[width=\columnwidth,height=!,keepaspectratio]{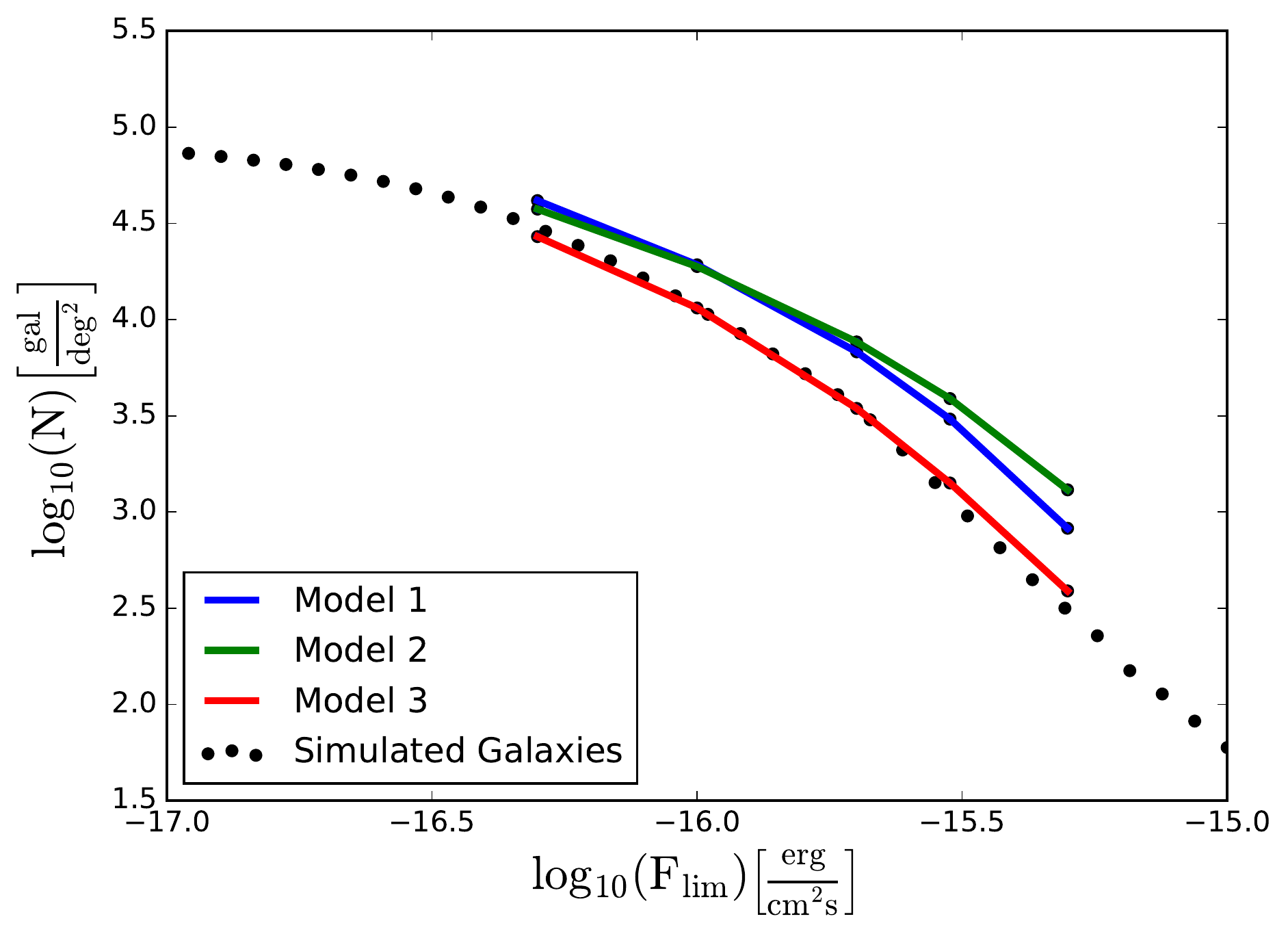}
\caption{Comparing the luminosity function of the \buzzard\ mock catalog to that predicted by \citet{pozzetti2016}. These are semi-analytic models made by fitting to observed luminosity functions from H$\alpha$ surveys. Specifically, Models 1 and 2 feature a Schechter parameterization, while Model 3 was designed specifically for high-redshift slitless surveys such as \WFIRST\ and Euclid, as it was fit directly to luminosity function data. This graphic is from one 54 deg$^2$ section in the sky; the redshift range is from $z=0.7$ to $z=1.5$ (similar to our Z2 bins, but matching the exact range used for the \protect\cite{pozzetti2016} model). 
}
\label{fig:luminosity}
\end{figure}

Galaxies are added to the simulation using the \textsc{ADDGALS} algorithm \citep{Wechsler2019}. Assuming an input luminosity function, this algorithm uses a model for density given absolute magnitude, $p(\delta|M_{r}, z)$ measured from a subhalo abundance matching (SHAM) model run on a smaller, higher resolution simulation. This model is then applied to the lower resolution lightcone simulations by drawing magnitudes from the assumed luminosity function, drawing densities from $p(\delta|M_{r},z)$, and assigning the galaxy to a particle in the lightcone with the correct density. After all rest frame $r$-band magnitudes are assigned to all galaxies, SEDs are then assigned from SDSS using the $\delta_g-M_{r}-$SED relation from SDSS \citep{Cooper06}, where $\delta_g$ is the distance to the fifth-nearest neighbor galaxy. The SEDs are represented by {\tt kcorrect} templates \citep{blanton2003} from which line fluxes and stellar masses can be determined. In Fig. \ref{fig:luminosity} we compare the \buzzard\ luminosity function to several empirical models based on grism and narrow-band data \citep{pozzetti2016}. To test the stellar mass function, we take the objects at flux $>1\times 10^{-16}\,$erg/cm$^2$/s, and compare the stellar masses from the catalog, versus using the flux $\rightarrow$ star formation rate (SFR) $\rightarrow$ stellar mass conversion assuming 1 mag extinction at H$\alpha$ and using the conversion factor for the \citet{2003PASP..115..763C} initial mass function (IMF) and the SFR sequence of Table 1 of \citet{2010MNRAS.405.1690D}. We obtain that at $z=1.0\pm0.2$ (1.9$\pm$0.2), the median stellar mass from the catalog is 0.23 (0.31) dex above the scaling relation, which is acceptable given the simplicity of the comparison.

\subsection{Assignment of the [N{\,\sc ii}]/H$\alpha$ ratio}

Here we outline our method to generate a [N{\,\sc ii}]/H$\alpha$ ratio for each galaxy. The [N{\,\sc ii}]/H$\alpha$ ratio is an observable in high-resolution spectra and is a commonly used metallicity indicator (see, e.g., \citealt{2002ApJS..142...35K, 2002MNRAS.330...69D, 2004MNRAS.348L..59P}), and so in principle one can use the metallicity to predict the [N{\,\sc ii}]/H$\alpha$ ratio. The full picture is more complicated: the Baldwin-Phillips-Terlevich (BPT) diagram of [O{\,\sc iii}]/H$\beta$ vs.\ [N{\,\sc ii}]/H$\alpha$ \citep{baldwin1981} has a sequnece of star-forming galaxies, ranging from low metallicity/high ionization (upper left) to high metallicity/low ionization parameter (lower right). The star-forming sequence evolves with redshift, which has been attributed to the N/O ratio varying at fixed metallicity (and instead being more closely correlated with stellar mass; e.g., \citealt{2016ApJ...828...18M}), or to massive binary stellar populations \citep[\eg][]{2016ApJ...826..159S}. We will circumvent this issue for the purpose of this paper by using an empirical mass-metallicity (MZ) relation with [N{\,\sc ii}]/H$\alpha$-based metallicities: uncertainties in interpretation of the line ratio cancel out when we predict [N{\,\sc ii}]/H$\alpha$ so long as we use the same calibration as in the MZ determination. Creating an accurate representation of this distribution within our simulation -- including the effects of environment -- is critical to the goals of this paper. Furthermore, the [N{\,\sc ii}]/H$\alpha$ ratio is expected to vary as a function of redshift \citep{kewley2013,masters2014,steidel2014,shapley2015,strom2017,kashino2017} similarly to the general MZ relation \citep{savaglio2005,maiolino2008,lilly2013}.

For each galaxy in the \buzzard\ mock catalog, [N{\,\sc ii}]/H$\alpha$ line strength ratios were assigned based on the stellar mass and redshift of that galaxy; we assume no other environmental trend in the mass-metallicity relationship itself. Specifically, we incorporate only the {\em mean} trend in the mass-metallicity correlation, without including effects due to the scatter in this relationship. We expect this to comprise the dominant systematic offset \citep[\eg][]{1979A&A....80..155L}, although including the scatter in the mass-metallicity relationship, which could conceivably be correlated with the galaxy environment \citep[\eg][]{2008MNRAS.390..245C}, may be an interesting investigation in future work. The dependence of the galaxy metallicity on star-formation rate is accounted for through the correlation between star-formation rate and redshift. This is justified observationally by \cite{wuyts2016}, who used the KMOS near-infrared multi-integral field unit survey to find the [N{\,\sc ii}]/H$\alpha$ ratio for 419 star-forming galaxies, in the redshift range $0.6 < z < 2.7$. They find that there is no significant dependence of the ratio on star-formation rate, given fixed redshift and stellar mass, although several studies have indicated possible connections between the line ratio and the specific star formation rate (SFR) \citep[e.g.][]{2010MNRAS.408.2115M,lilly2013,2015A&A...577A..14M,2016ApJ...828...18M,faisst2017}. Furthermore, this assumption has support from hydrodynamical simulations; \cite{hirschmann2017} found that the primary evolutionary trends were based on redshift and stellar mass, in a manner consistent with our model. Other effects such as specifics of their AGN model and ionized-gas hydrogen/electron density were found to have a much smaller impact on the cosmic evolution than the galaxy mass.

\begin{figure}
\centering
\includegraphics[width=\columnwidth,height=!,keepaspectratio]{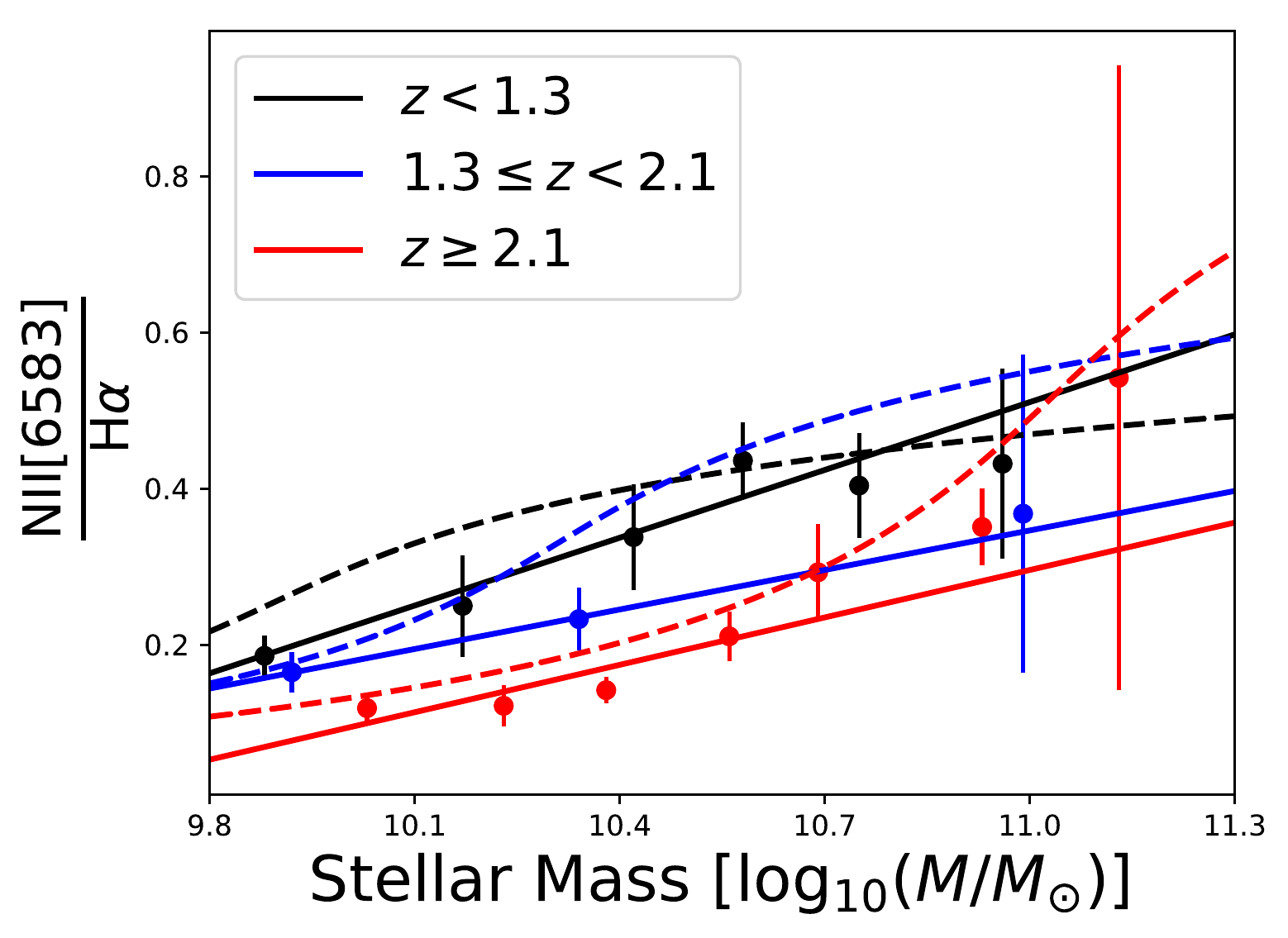}
\caption{Empirical relationship (solid lines) between [N{\,\sc ii}]/H$\alpha$ ratio and log stellar mass. Data being fitted was obtained from \citet{wuyts2016}. These relationships were used to construct the relative [N{\,\sc ii}] line strength within the mock galaxy sample. For comparison, we also plot in dashed lines the fits at redshifts 0.9 (black), 1.5 (blue), 2.3 (red), from Eq.~(3) in \citet{faisst2017}. Given the large uncertainties in the data, our fits for redshift larger than 1.3 may or may not underestimate the line ratio. Note that both references adopt the \citet{2003PASP..115..763C} IMF, consistent with the {\tt kcorrect} templates in our simulation.}
\label{fig:MZR}
\end{figure}

We base our fits on the results from \cite{wuyts2016}, who grouped their data into galaxy mass sub-ranges spanning $\mathrm{log_{10}(M/M_{\odot})} =9.88- 11.13$, for redshift bins at $z \approx 0.9$, $z \approx 1.5$, and $z \approx 2.3$. We used these empirical relationships to create three linear fits, one for each redshift bin, giving the [N{\,\sc ii}]/H$\alpha$ ratio as a function of log stellar mass. We then used these fits to provide our catalog with nitrogen line strength ratios, depending on the mass and redshift of each galaxy in our sample. The fits can be seen in Fig.~\ref{fig:MZR}.

In our analysis, we make no special allowance for AGNs. We use the ``All'' fits from \cite{wuyts2016}, so AGN line ratios do pull the fits; however sample sizes in some bins are small, the selection effects may be different from \WFIRST, and the effect on higher-order moments of the [N\,{\sc ii}]/H$\alpha$ distribution (scatter, correlation with other galaxy properties, etc.) are not captured by this procedure, and may be revisited in future work. Furthermore, our analysis is only valid for galaxies on the star-forming main-sequence and not for starbursts, whose [N\,{\sc ii}]/H$\alpha$ ratio could be substantially different due to their high specific SFR with respect to main-sequence galaxies at the same stellar mass and redshift.

Approximately 0.1\% of objects in the mock catalog have very small stellar masses, which when combined with the linear fit from Fig. \ref{fig:MZR} produce a negative [N{\,\sc ii}]/H$\alpha$ ratio. These ratios are set to zero. This happens for galaxies with $\mathrm{log_{10}(M/M_{\odot})} < (9.2, 8.9, 9.5)$ for $z \approx (0.9, 1.5, 2.3)$, respectively. We do not expect this simplification to have any effect on our full sample results, due to the relatively small number of galaxies affected\footnote{Most of those low-mass galaxies will not pass the WFIRST flux limit ($\sim 10^{-16}\,$erg/s/cm$^2$). Considering a galaxy at the lowest redshift (0.705) and with the limiting flux, with the fiducial cosmology, we estimate the line luminosity to be $2.2\times 10^{41}\,$erg/s. The H$\alpha$ luminosity is lower than that, roughly $1.38\times 10^{41}\,$erg/s if we take the line ratio as 0.6. This provides a conservative estimate for the H$\alpha$ luminosity, hence the SFR, which is given by SFR$/[M_\odot\,{\rm yr}^{-1}]\approx 0.57\times 7.9\times 10^{-42} L_{{\rm H}\alpha}$/[erg/s]$\sim 0.62$ \citep[see Eq.~10.109 in][]{2010gfe..book.....M}, where 0.57 is the conversion from Salpeter to Chabrier IMF \citep{2010ApJ...725..742M}. Using the SFR-stellar mass relation for the main-sequence galaxies \citep[e.g.][]{2010MNRAS.405.1690D}, we find stellar mass to be around $3\times 10^{10}\,M_\odot$, still above the masses where the ``negative ratio'' is concerned.}, and because fitting these lower-mass galaxies with a more complex model would likely still result in [N{\,\sc ii}]/H$\alpha$ near zero.

\subsection{Data binning and redshift distributions}
\label{subsec:binning}

In order to generalize our analysis to many possible future surveys with a range of redshifts and flux thresholds, we binned our simulation catalog by both redshift and flux, and computed the correlations for all galaxies within each bin. The redshift bins are differential, while the flux bins are integral; their ranges are listed in Table \ref{tab:binning2}. We measure galaxy flux by the {\em total} emission of the combined [N{\,\sc ii}] and $H\alpha$ lines, which is consistent with the future measurements that will be made by \WFIRST. In order to decrease computation time, each bin was further divided into six congruent geometric regions on the sky (S1 to S6, displayed in Fig. \ref{fig:regions}). Correlation functions were independently generated using the counts from each sector.

\begin{table}
\centering 
\caption{\label{tab:binning2}The number of galaxies, in millions, within each bin. This is the sum total of galaxies within each geometric region for the data counts. The galaxy H$\alpha$+[N{\,\sc ii}] flux $F$ is measured in units of $10^{-17}\, \mathrm{erg}\,\mathrm{cm}^{-2}\,\mathrm{s}^{-1}$. In parenthesis is the label we use to reference each bin throughout the text. Note that the WFIRST flux limit for point sources is $F=5$ \citep{spergel2015}, and around 10 for galaxies.}
\begin{tabular}{c c c c} 
\hline\hline 
 &  $F>$ 8 &  $F>$ 13 & $F>$ 25 \\ [0.5ex] 
\hline 
0.705 $\leq z <$ 1.345 & - & 65.1 (Z2H2) & 17.8 (Z2H3)  \\
1.355 $\leq z <$ 1.994 & 59.1 (Z3H1) & 24.9 (Z3H2) & 5.2 (Z3H3)  \\[0.5ex]
\hline 
\end{tabular}
\end{table}

\begin{table}
\centering 
\begin{tabular}{c c} 
\hline\hline 
Bin & Median Redshift \\ [0.5ex] 
\hline 
Z2H2 & 0.9116 \\
Z2H3 & 0.8698 \\
Z3H1 & 1.6136 \\
Z3H2 & 1.6134 \\
Z3H3 & 1.6169 \\ [0.5ex]\hline 
\end{tabular}
\caption{The median redshift of each bin.}
\label{tab:redshifts} 
\end{table}

To ensure that the correlation functions for different groups of measurements, i.e. $z_{\mathrm{obs}}$ and $z_{\mathrm{true}}$, can be compared within the exact same redshift limits, it is necessary to use slightly different subsets of galaxies for different calculations -- for example, one specific galaxy whose true redshift falls in redshift bin 1, may have its observed redshift place it within bin 2. Because of this, the exact galaxy samples vary slightly between true and observed samples.

Each galaxy was given a set of redshifts: a true redshift, an observed redshift, and a shuffled redshift. The true redshift, $z_{\mathrm{true}}$, is simply the original redshift value from the catalog -- that is, the redshift that would be observed if the [N{\,\sc ii}] and H$\alpha$ lines could be separately resolved. This redshift still incorporates the peculiar velocities of each galaxy in the redshift value.

We use the observed redshift value, $z_{\mathrm{obs}}$, to include two separate effects. First, we insert the statistical error in the wavelength centroid. Each galaxy's observed redshift is modified by adding to it a number generated from a Gaussian with mean zero, and standard deviation:
\begin{equation} \label{eq:obs-spread}
\sigma_{z} = 10^{-3}(1 + z).
\end{equation}
This error is statistical, and thus not dependent on each individual galaxy's metallicity. This is the error specified by the \WFIRST\ SRD. The real errors will also depend on line flux and galaxy size ($\sigma_z$ is smaller for galaxies that have brighter lines or smaller angular sizes), however assessment of this is outside the scope of this paper.\footnote{Studying how the statistical variance $\sigma_z^2$ depends on galaxy properties will involve both the grism image simulations and (ultimately) \WFIRST\ deep field data, which will empirically constrain the precision of the redshift measurement by repeating it many times for the same sample of galaxies.}

The second effect incorporated into $z_{\mathrm{obs}}$ is the [N{\,\sc ii}]+H$\alpha$ line blending effect. As long as the nitrogen line strength is non-zero, this will pull each galaxy's redshift toward a ``redder'' value, due to the anisotropy of the nitrogen line pair. This effect is calculated by finding the offset from the H$\alpha$ line center, in the rest frame of the galaxy:
\begin{equation}
\delta \lambda = \frac{\Delta_{1}F_{6585} - \Delta_{2}F_{6550}}{F_{{\rm H}\alpha} + F_{6585} + F_{6550}}.
\label{eq:delta-lambda}
\end{equation}
Here $\delta \lambda$ is the offset of the observed line from the true H$\alpha$ line, $F_{6550}$ and $F_{6585}$ refer to the strengths of the respective [N{\sc\,ii}] lines, $F_{{\rm H}\alpha}$ is the strength of the H$\alpha$ line, $\Delta_{1}$ is the difference in wavelength between H$\alpha$ and [N{\,\sc ii}] 6585, and $\Delta_{2}$ is the difference in wavelength between [N{\,\sc ii}] 6550 and H$\alpha$. (Both $\Delta_1$ and $\Delta_2$ are defined to be positive.) We use the vacuum values of 6549.86 \AA\ and 6585.27 \AA\ for the ($^1$D$_2^{\rm e}-^3$P$_1^{\rm e}$) and ($^1$D$_2^{\rm e}-^3$P$_2^{\rm e}$) transitions of N{\,\sc ii}, respectively, and we use a value of H$\alpha$ at 6564.61 \AA. We take the ratio in strengths between the two nitrogen lines, [N{\,\sc ii}] 6550/[N{\,\sc ii}] 6585 to be 0.32567 \citep{storey2000}. Note that Eq.~(\ref{eq:delta-lambda}) is valid in the extreme case of a completely unresolved line; the marginally resolved case (relevant for \WFIRST) can lead to smaller shifts \citep{faisst2017}, and hence our analysis is conservative.

Once the offset for a specific galaxy is calculated, we can find the observed redshift using
\begin{equation}
z_{\rm obs} = z_{\rm true} + \frac{\delta \lambda}{\lambda_{{\rm H}\alpha}}(1 + z_{\rm true}) + \delta_{\rm st},
\end{equation}
where $\delta_{\rm st}$ is a realization of the statistical error (see Eq. \ref{eq:obs-spread}). Most galaxies show differences between the true redshift and the observed redshift at or below $\delta z = 10^{-3}$, due to the natural [N{\sc\,ii}]/H$\alpha$ ratio empirically found in galaxies. There is a small subset of galaxies, of order 0.1\%, with no difference between observed and true redshift, due to the linear metallicity fit. However, the majority of this subset is eliminated during the binning process, as we remove galaxies below a certain flux threshold, and low flux is correlated with the low stellar mass used in metallicity fitting.

\begin{figure}
\centering
\includegraphics[width=\columnwidth,height=!,keepaspectratio]{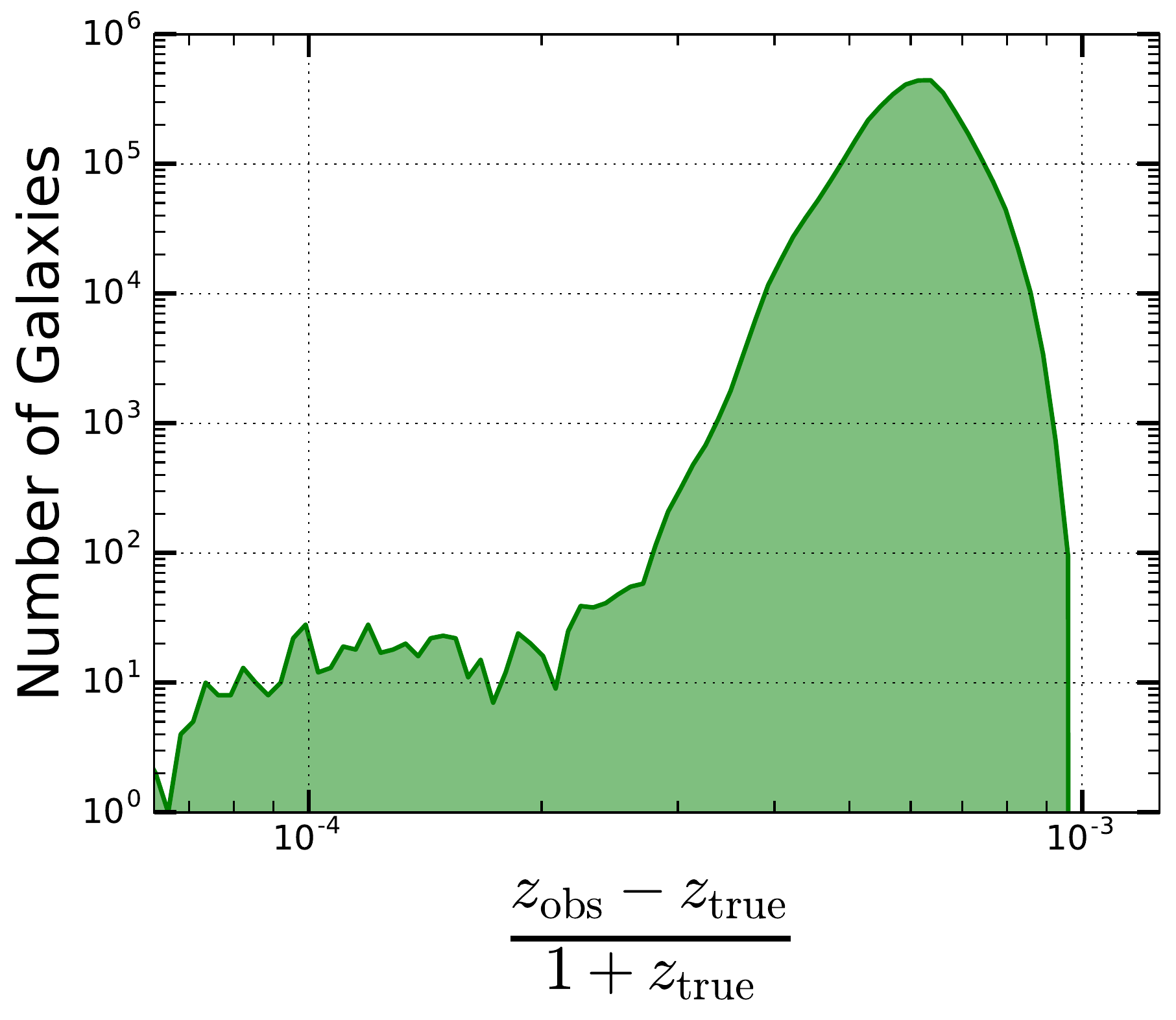}
\caption{The difference between the observed redshift and the true redshift for a sample of a little over 10 million galaxies. We included only the mean trend in the [N{\,\sc ii}]/H${\alpha}$ ratio in the mass-metallicity relationship, but not the scatter. (We expect the mean trend to capture the lowest-order correlation between line ratio and environment, but of course excluding the scatter results in an artificially narrow peak.) This does not include the Gaussian photon noise smearing, only the difference due to the line blending effect. This histogram is generated from one sky section of the parent \buzzard\ sample, cut from $z=0.7$ to $z=1.3$ (the approximate redshift range of the Z2 bins), and covers approximately 54 square degrees of sky.}
\label{fig:zdiff}
\end{figure}

\begin{figure*}
\centering
\includegraphics[scale=0.75]{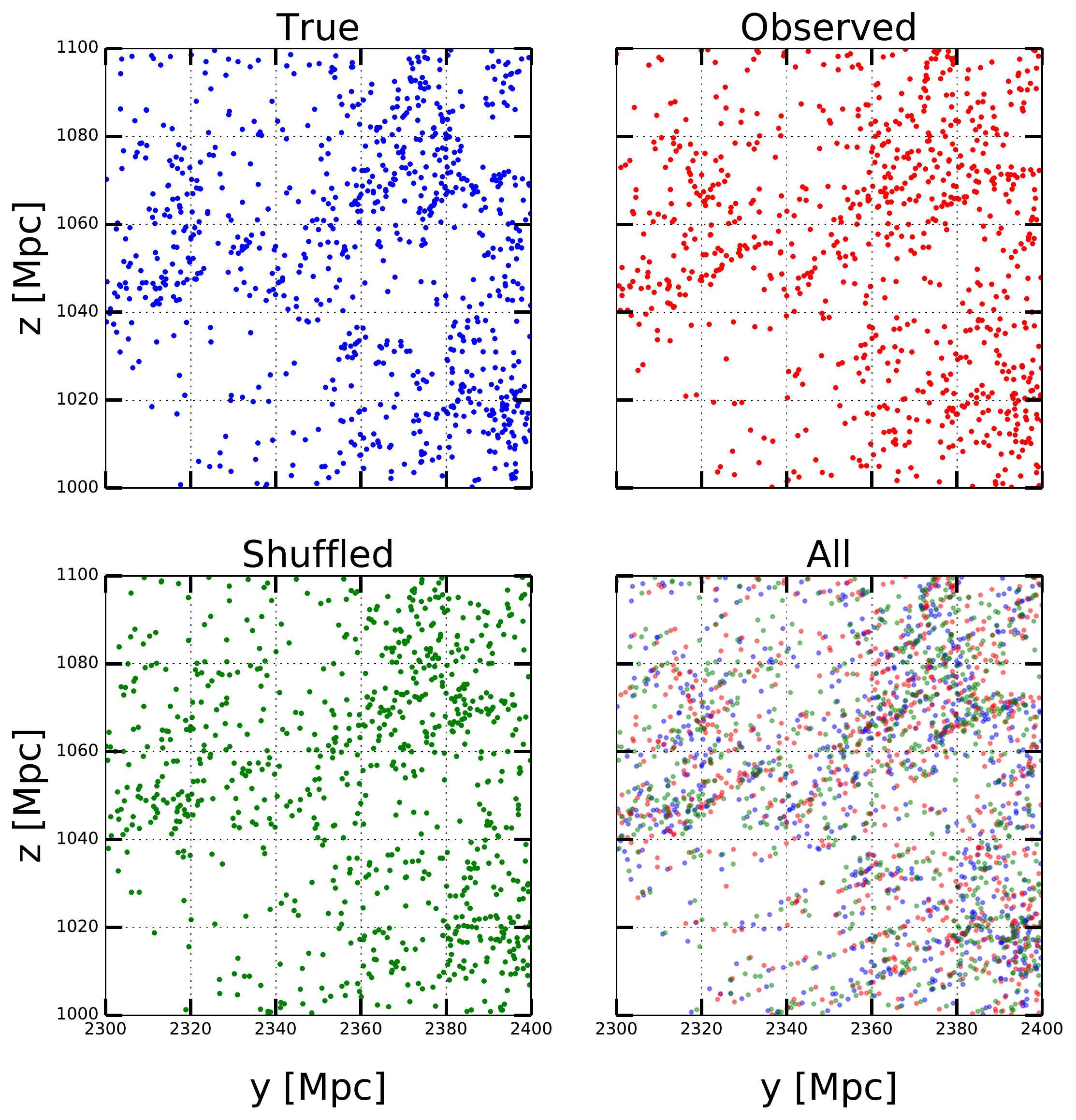}
\caption{We display the physical positions of the galaxies given the redshifts of each of our true, observed, and shuffled catalogs. These images were generated using Regions 2 and 5 of the Z2H3 catalogs. We have selected all galaxies with a Cartesian x-coordinate (defined by $x=D_c \cos({\rm DEC})\cos({\rm RA})$, where $D_c$ is the comoving distance) between -50 and 50 Mpc, a z-coordinate (defined by $z=D_c\sin({\rm RA})$) between 1000 and 1100 Mpc, and a y-coordinate (defined by $y=D_c \cos({\rm DEC})\sin({\rm RA})$) between 2300 and 2400 Mpc. The galaxies in the x-direction have been projected into the y-z plane, and the observer is located at the origin, toward the bottom-left corner of the image (which can be seen by the `streaks' due to the changes in redshift for the combination of the catalogs).}
\label{fig:data_comparison}
\end{figure*}

After each galaxy has values for $z_{\mathrm{true}}$ and $z_{\mathrm{obs}}$, we can then generate the shuffled redshift, $z_{\mathrm{shf}}$. This is done by creating a list of $\delta z$ values:
\begin{equation} \label{eq:deltaz}
\delta z \equiv \frac{z_{\rm obs} - z_{\rm true}}{1+z_{\rm true}}.
\end{equation}
We then shuffle the list, matching each $\delta z$ with a true galaxy to create a ``shuffled'' redshift:
\begin{equation}
z_{\rm shf} \equiv (1+z_{\rm true}) \delta z + z_{\rm true}.
\end{equation}
This creates a galaxy catalog where the redshift error distribution is identical to that in the observed catalog, but where all correlations between $\delta z$ and galaxy environment are destroyed. In this way, by comparing results from the observed distribution to the shuffled distribution, we can see whether parameter offsets are due to the distribution of redshift errors, or correlations between the redshift error distribution and galaxy environment. This will have an important impact in how we approach mitigation. In Fig. \ref{fig:data_comparison} we display the differences in inferred position for a subset of galaxies within the catalog.

\subsection{Correlation functions}

Here, we detail our pipeline for calculating the correlation functions within each ZH bin. The pipeline uses the same code as that described in \cite{martens2018}. 

\begin{figure*}
\centering
	\begin{tabular}{@{}cc@{}}
  	\includegraphics[width=\columnwidth]{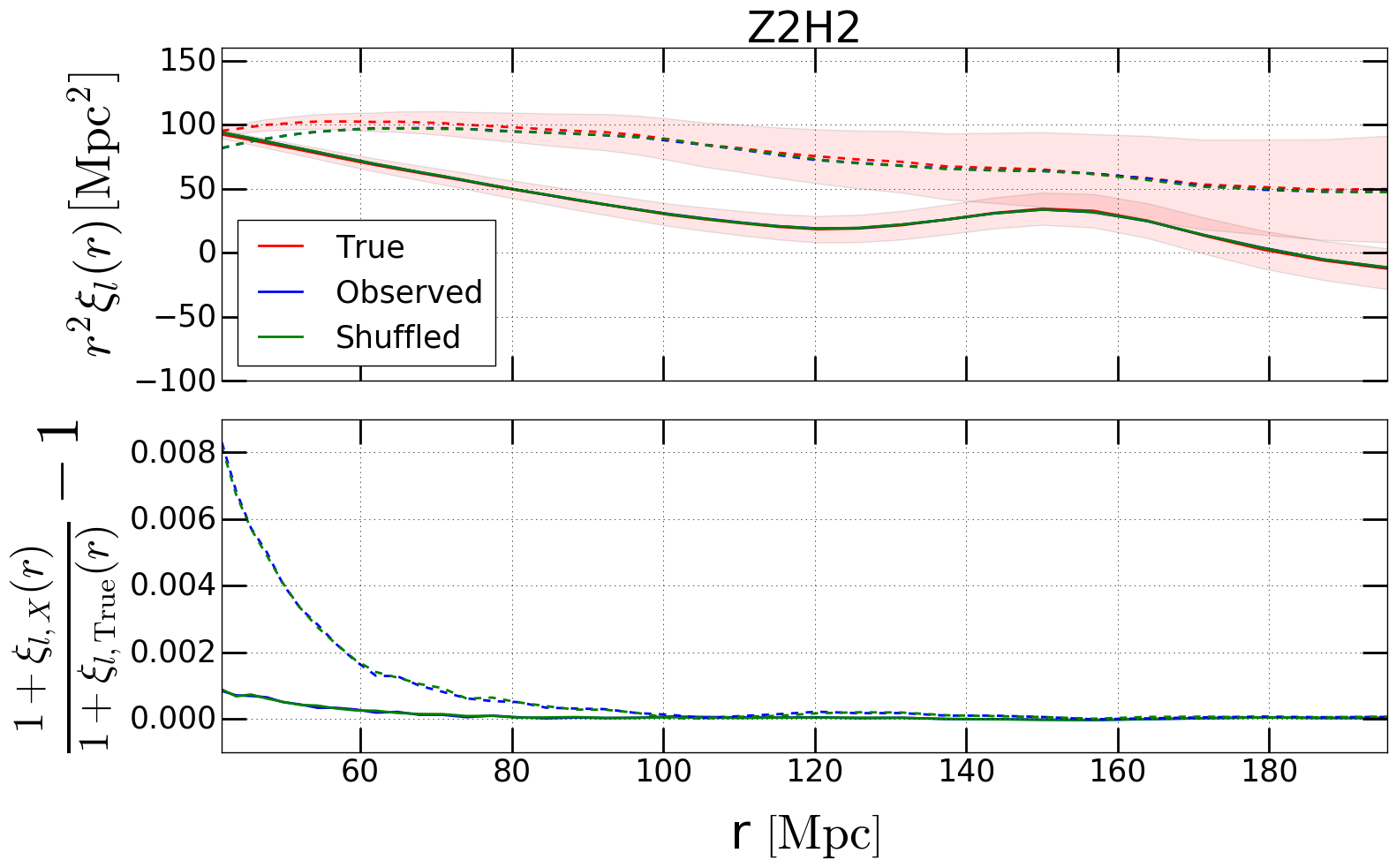} &
    \includegraphics[width=\columnwidth]{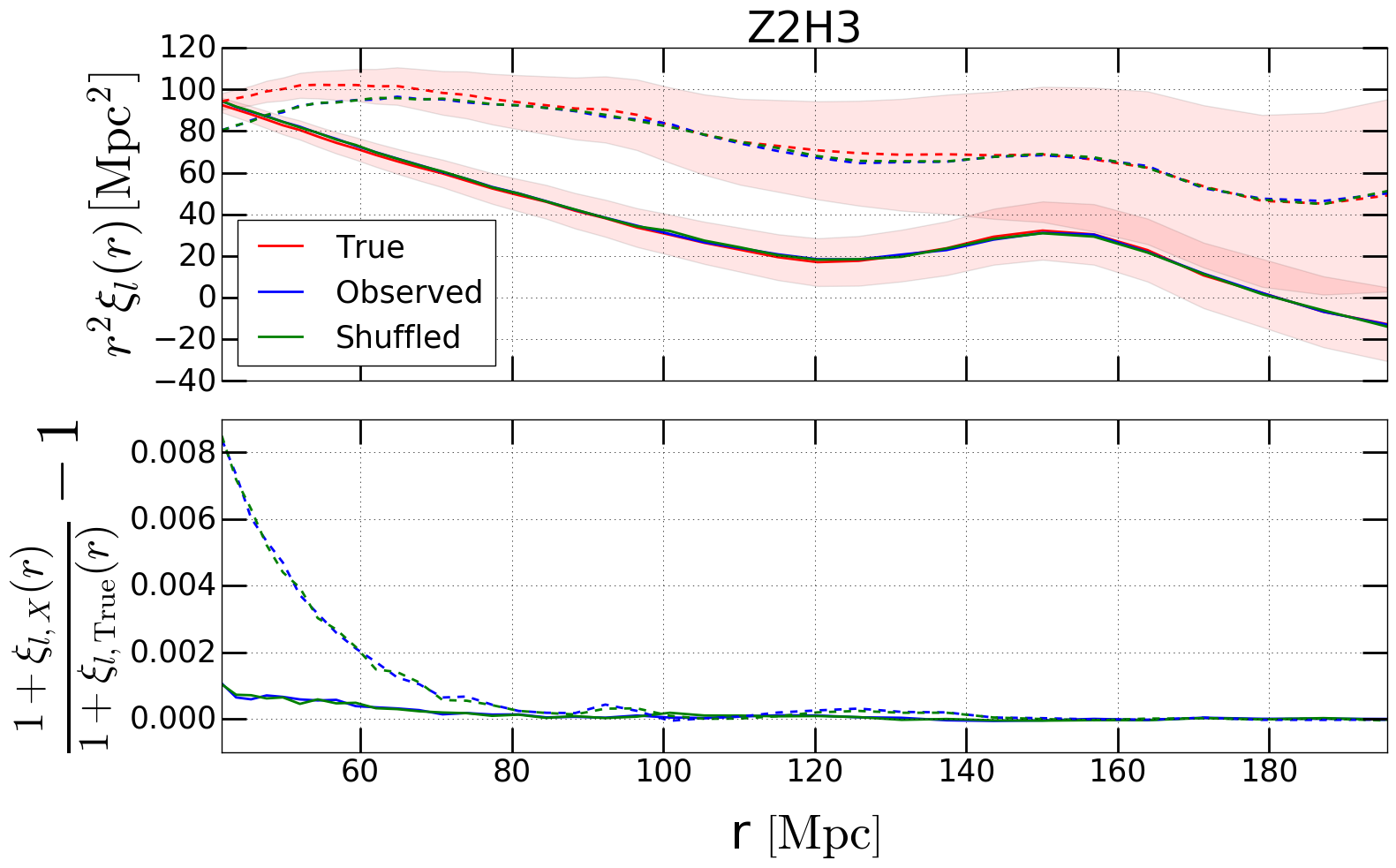} \\
    \includegraphics[width=\columnwidth]{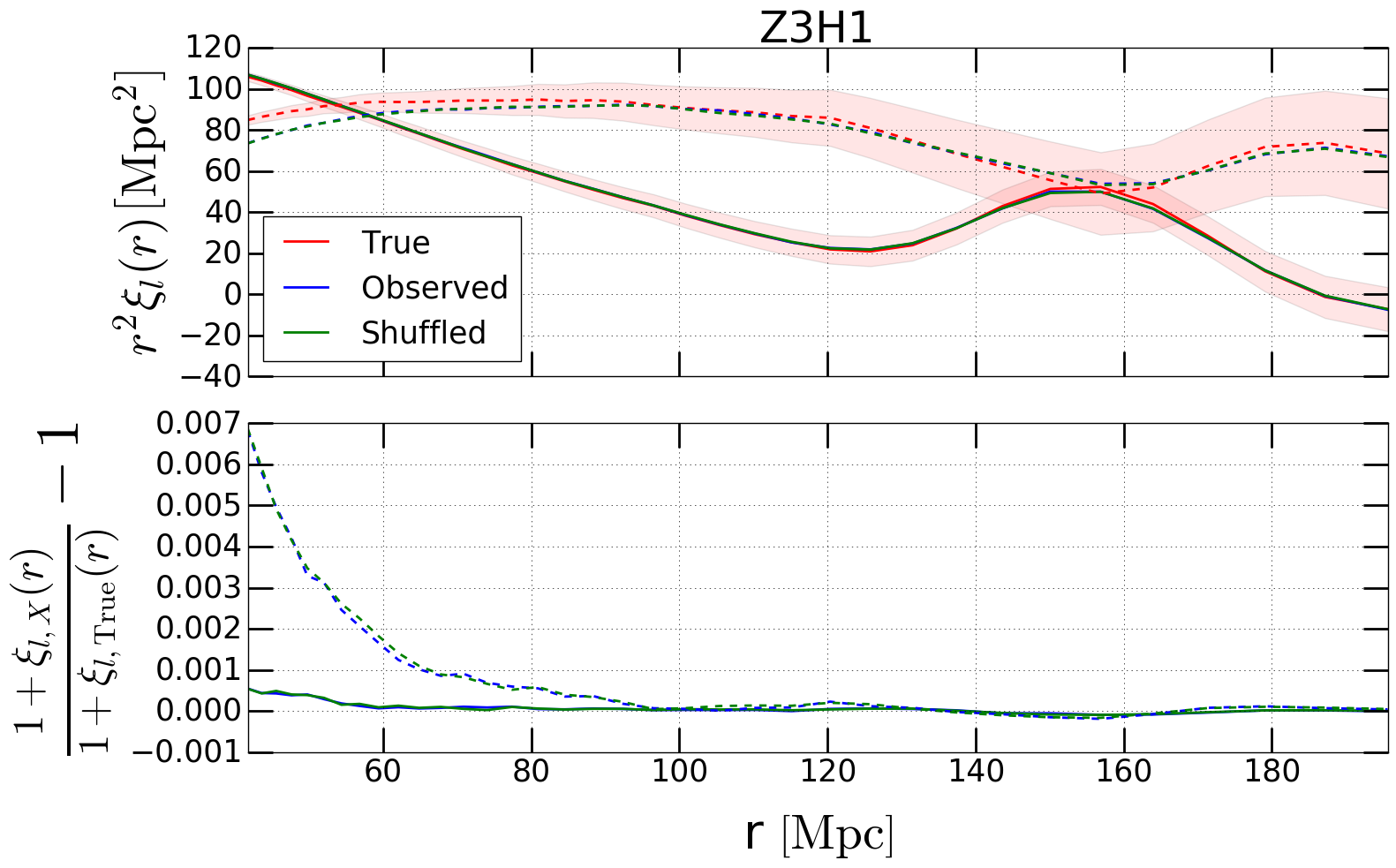} &
    \includegraphics[width=\columnwidth]{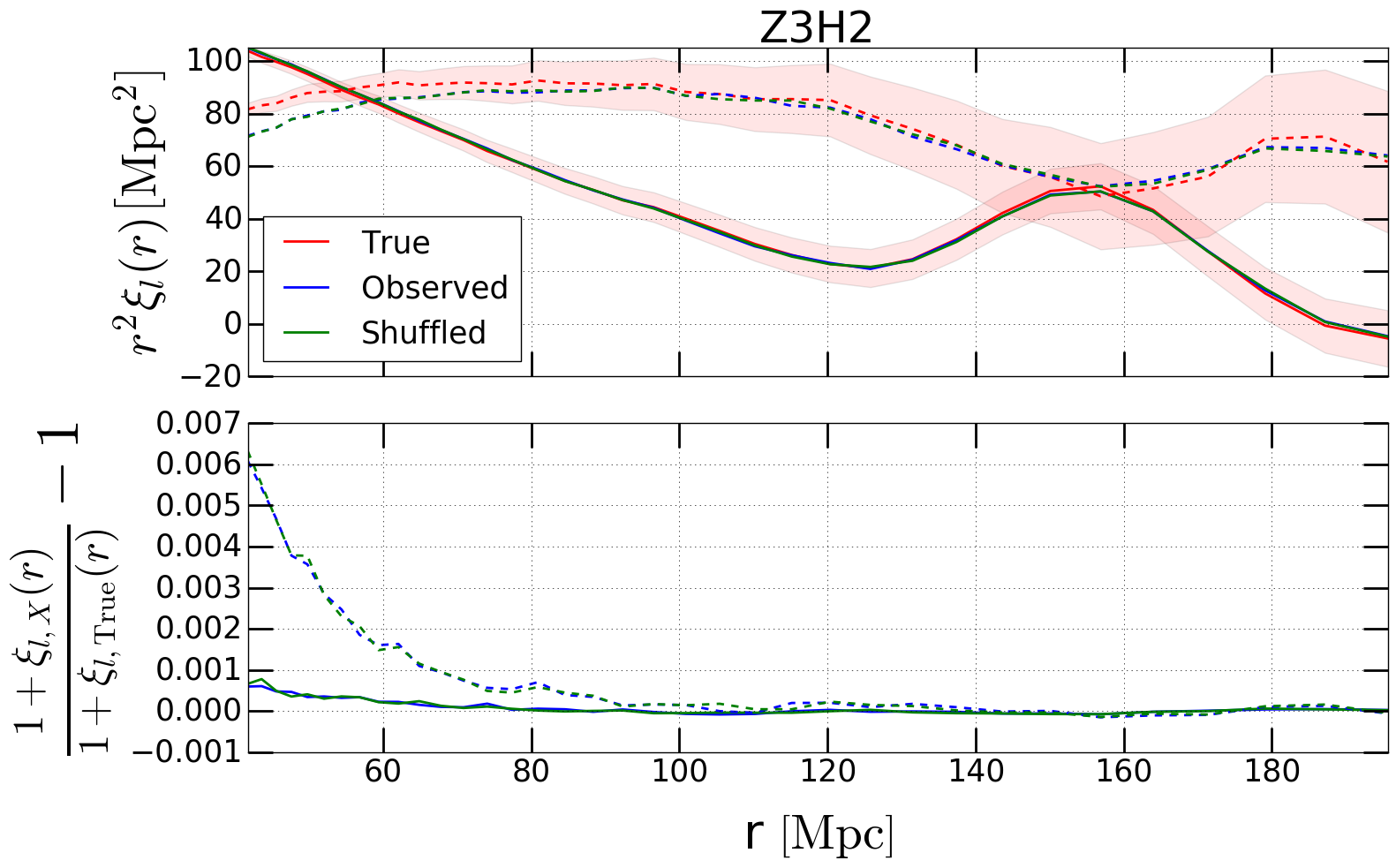} \\
    \includegraphics[width=\columnwidth]{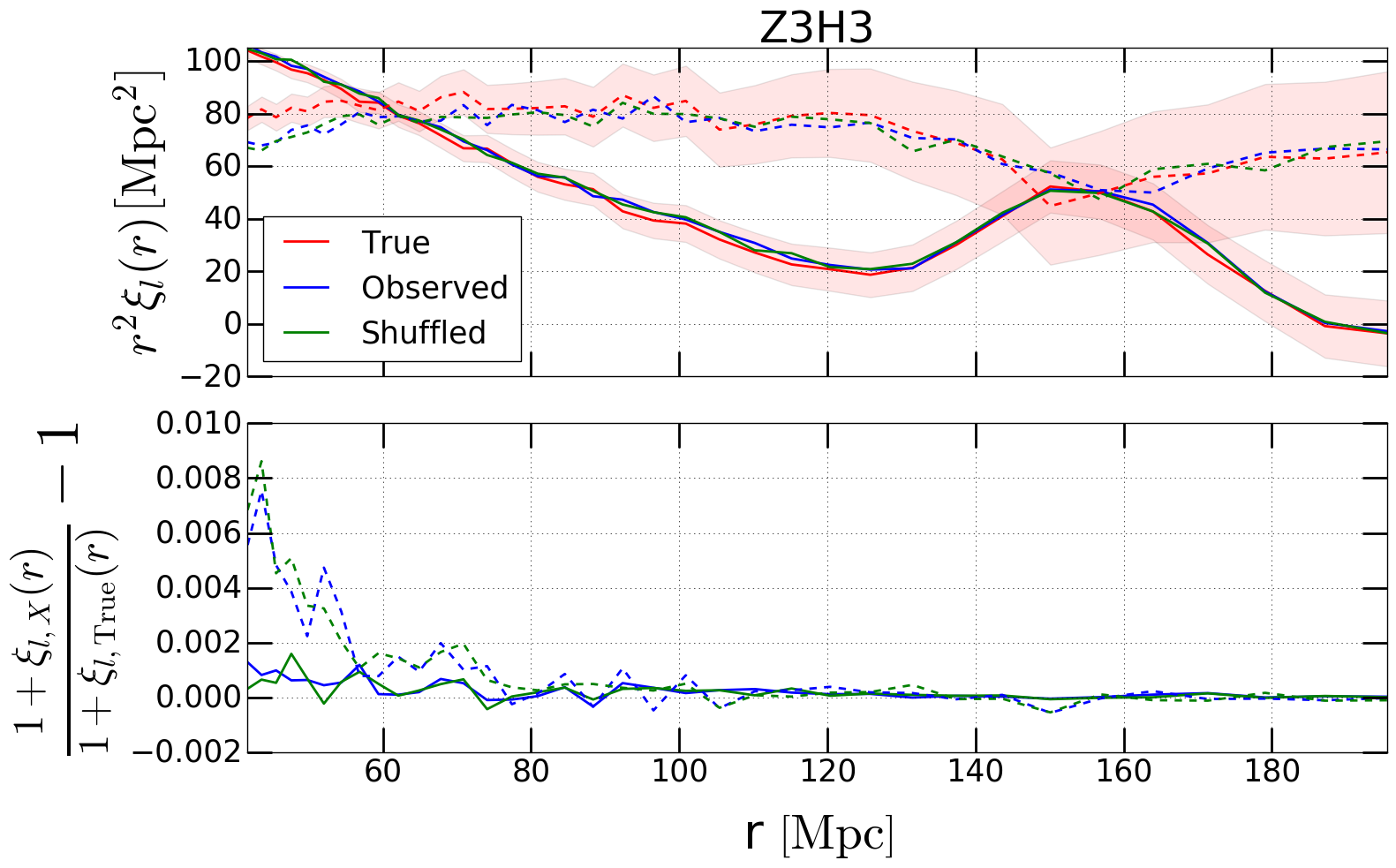} &
    \includegraphics[width=\columnwidth]{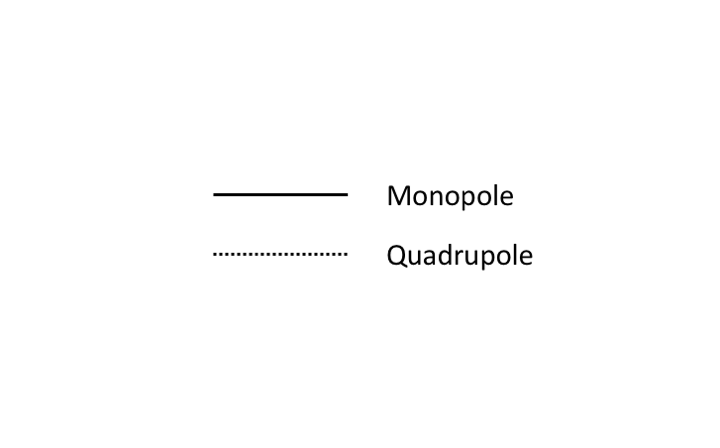} \\
  	\end{tabular}
\caption{For each ZH bin, we compare the monopole and quadrupole correlation functions for each of the true, observed, and shuffled redshift catalogs. The top panel of each plot displays the correlation function monopoles (solid lines) and quadrupoles (dashed lines). Both quadrupoles and monopoles are multipilied by $r^{2}$, and the quadrupoles are multiplied by $-1$. The bottom panel of each plot shows the fractional comparison between the observed and shuffled correlation multipoles, as a function of separation $r$.}
\label{fig:corr_comp}
\end{figure*}

Once pair counts were obtained for the true, observed, and shuffled samples within each flux and redshift bin, the redshift distributions of each were used to generate random catalogs. Randomly placed galaxies were created and given redshifts pulled from the distribution of the matching data bin. The random galaxy count is equal to 3 times the simulated real galaxy count, which was decided on with the intent to minimize error associated with random galaxy shot noise, but also work within the limits of computation time.\footnote{For the final \WFIRST\ analysis, more resources will be available to devote to random pair counting. The random catalog shot noise is a fraction $n_{\rm D}/n_{\rm R}$ of the shot noise in the data, so to make this negligible, we will need a $n_{\rm R}/n_{\rm D}$ much greater than that used in this paper.} Pair counts were done on the random catalogs in order to construct the correlation function using the Landy-Szalay method \citep[LS:][]{landy1993,peebles1974}. The correlations are calculated as a function of redshift-space separation $s$ and $\mu = \cos\theta$, where $\theta$ is the angle with respect to the line of sight. The LS estimator is 
\begin{equation}
\xi(s,\mu) = \frac{\mathrm{DD}(s,\mu) - 2\mathrm{DR}(s,\mu) + \mathrm{RR(s,\mu)}}{\mathrm{RR}(s,\mu)},
\end{equation}
where DD refers to the number of pairs of galaxies within a specific distance shell, $s + \Delta s/2$ and $s - \Delta s/2$, and within a specific angular range $\mu + \Delta \mu/2$ and $\mu - \Delta \mu/2$, for the data sample. RR refers to the same, but for the random sample. DR refers to a combined catalog of data and randoms, where we are counting pairs of opposite types only. We use 120 logarithmically spaced radial bins, from $s = 1$ Mpc to $s = 200$ Mpc. Both DD and RR counts are normalized by the total number of galaxies in that bin, specifically:
\begin{equation}
\mathrm{DD \rightarrow \frac{DD}{n_{D}(n_{D}-1)}}~~~{\rm and}~~~ \mathrm{RR \rightarrow \frac{RR}{n_{R}(n_{R}-1)}},
\end{equation}
while the DR counts are normalized by:
\begin{equation}
\mathrm{DR \rightarrow \frac{DR}{n_{D}n_{R}}}.
\end{equation}
Our correlation function code calculates pairs in 20 $\mu$-bins, from $-1$ to $+1$, with a separation of $\Delta \mu = 0.1$. Although the simulations were generated in $\mu$-binned ``wedge'' space, we convert them to multipole space for parameter fitting. The formula for conversion is the same as that used in SDSS BOSS analysis \citep{ross2016}:
\begin{equation}
\xi_{l}(r) = \frac{2l + 1}{2}\sum_{i=1}^{i_{\rm max}}\frac{1}{i_{\rm max}}\xi(r, \mu_{i})L_{l}(\mu_{i}),
\end{equation}
where $\mu_{i} = (i - 1/2)/i_{\rm max}$, $L_{l}$ is the Legendre polynomial\footnote{To avoid confusion with power spectra, we use $L$ instead of the traditional $P$ for Legendre polynomials.} of order $l$, $\xi$ is the $\mu$-binned correlation function, and $i_{\rm max}=20/2=10$ is the number of $\mu$-bins from 0 to $+1$. In Fig. \ref{fig:corr_comp}, we plot the resulting monopole and quadrupole correlation functions for the Z2H2, Z2H3, Z3H1, Z3H2, and Z3H3 bins, respectively. In each plot, we show the comparison of the quadrupole and monopole correlation functions between the true, observed, and shuffled redshift catalogs. We also show the fractional comparison from the true catalog to the observed and shuffled catalogs.

\section{Parameter fitting}
\label{sec:fitting}

In this section, we detail the process of fitting our correlation functions from the simulation with smaller scale RSD parameters and large scale BAO parameters. We perform the RSD and BAO fits completely independently, and in separate scale ranges; although in projects whose primary goal is parameter measurement we would fit both parameter sets together, here we separate them to provide greater sensitivity to the systematic errors we are studying, resulting in the most conservative choice for systematic error budgeting.

First, we use Convolution Lagrangian Perturbation Theory \citep[CLPT,][]{carlson2012} and Gaussian Streaming Redshift Space Distortions \citep[GSRSD,][]{wang2014} to fit the parameters $b_{g}$, the linear galaxy bias, and $f_{v}$, the dimensionless linear growth factor, on scales of 42 to 200 Mpc (for varying values of $\sigma_{\rm FOG}^{2}$, the variance of small-scale dispersion due to FOG, the Finger-of-God effect, in the unit of length$^2$). This fit focuses on small scale clustering parameters, and is explained in detail in Sec. \ref{subsec:rsd_parameter}. Second, we fit BAO parameters on scales of 60 to 200 Mpc, described in Sec. \ref{subsec:BAO}. In both cases, we use the correlation function monopole and quadrupole to drive the fitting function.\footnote{The scale ranges are chosen for several reasons: (1) the covariance matrix we use is from \cite{grieb2016}, which was tested on scales 30-180$\,h^{-1}$Mpc, \ie, 42-257$\,$Mpc, giving the lower scale cut for the RSD fits; (2) scales 60-200$\,$Mpc is sufficient to cover the BAO peak and the broadband feature while not being affected by small-scale RSD effects; and (3) cutting the scales at 200$\,$Mpc not only reduces the computation time but also avoids the edge effects due to the finite survey and region areas. We will see in Sections \ref{subsec:rsd_parameter} and \ref{subsec:BAO} that the final results are robust when testing with various scale cuts.}

\subsection{Covariance matrices}

To provide a best-fit to the observed correlation functions, we assume a Gaussian-distributed likelihood for our vector of measured correlation functions:
\begin{equation}
\mathcal{L}(\bmath{p}) \propto e^{- \chi^{2}(\bmath{p})/2},
\end{equation}
where $\chi^{2}$ is given by: 
\begin{equation} \label{eq:chi_squared}
\chi^{2}(\bmath p) = \sum_{\ell,\ell'}\sum_{i,j} (\xi^{i}_{\ell}(\bmath{p}) - \hat{\xi}^{i}_{\ell} ) [{\mathbfss C}^{-1}]_{\ell \ell' i j} (\xi^{j}_{\ell'}(\bmath{p}) - \hat{\xi}^{j}_{\ell'} ).
\end{equation}
Here $\bmath p$ is a vector of parameters; $\ell$ and $\ell'$ are the moments of the correlation function (here equal to 0 or 2); $i,j$ refer to the separation bins; $\hat{\xi}$ is the measured correlation function; $\xi$ is the model correlation function; and ${\mathbfss C}$ is the covariance matrix \citep{sanchez2008,cohn2006}, which we calculate using the method from \cite{grieb2016}. \cite{grieb2016} generate a theoretical model for the linear covariance of anisotropic galaxy clustering observations, making use of synthetic catalogs. As input, the calculation of the covariances are based on an input linear galaxy power spectrum dependent on both the wavevector and the angle with the line of sight, $P(k,\mu)$. In order to calculate this, we first calculate the linear matter power spectrum using CLASS \citep{blas2011} and then compute the no-wiggle power spectrum from the formulae listed in \cite{eisenstein1998}. This is done at the median redshift of each sample. We next account for redshift space distortion effects to the power spectrum using the procedure outlined in \cite{ross2016}:
\begin{equation}
P(k,\mu) = b^{2} C^{2}(k,\mu,\Sigma_{s})\left[(P_{\rm nonlin} - P_{\rm nw})e^{-k^{2}\sigma_{v}^{2}} + P_{\rm nw} \right]
\end{equation}
where the no-wiggle power spectrum is also generated with the nonlinear power spectrum from the HaloFit model (taking no-wiggle from \cite{eisenstein1998} as input)\citep{2003MNRAS.341.1311S}. We have used
\begin{equation}
\sigma_{v}^{2} = (1 - \mu^{2}) \Sigma_{\perp}^{2}/2 + \mu^{2}\Sigma_{\parallel}^{2}/2
\end{equation}
and
\begin{equation}
C(k,\mu,\Sigma_{s}) = \frac{1+\mu^{2}\beta}{1 + k^{2}\mu^{2}\Sigma_{s}^{2}/2}.
\end{equation}
We define the spreading due to photon noise:
\begin{equation}
\Sigma_{s,\mathrm{phot}} = (300\,{\rm km}\,{\rm s}^{-1})\times \frac{1+z}{H(z)},
\end{equation}
where we set $\Sigma_{s}^{2} = \Sigma_{s,\mathrm{phot}}^{2} + (2.26\,{\rm Mpc})^{2}$. For the True catalog, $\Sigma_{s,\mathrm{phot}}$ is set to 0. This effectively incorporates the spreading we have added in the shuffled and observed catalogs due to uncertainty in photon noise, and is displayed in Table \ref{tab:cov_vals}. 

\begin{table}
\centering 
\begin{tabular}{c c c c c} 
\hline\hline 
Bin & $\beta$ & $\Sigma_{s} (\mathrm{Mpc})$ & $\Sigma_{\perp} (\mathrm{Mpc})$ & $\Sigma_{\parallel} (\mathrm{Mpc})$ \\ [0.5ex] 
\hline 
Z2H2 & 0.5762 & 4.97 (5.46) & 7.422 & 13.709 \\
Z2H3 & 0.5800 & 4.99 (5.48) & 7.520 & 13.847\\
Z3H1 & 0.4382 & 4.64 (5.16) & 5.679 & 10.956\\
Z3H2 & 0.4424 & 4.64 (5.16) & 5.681 & 10.959\\
Z3H3 & 0.4362 & 4.64 (5.16) & 5.679 & 10.957\\ [0.5ex]\hline 
\end{tabular}
\caption{Parameters used in generating $P(k,\mu)$ for both the covariance matrix generation and the BAO models used for fitting. These are calculated using formulae from \protect\cite{seo2007}. For $\Sigma_{s}$, the first value indicates the value used for the True catalogs, while the second value was used for the Observed and Shuffled catalogs; it was increased due to the extra variance added to simulate photon noise.}
\label{tab:cov_vals} 
\end{table}

In these equations, our values for $\beta$, $\Sigma_{s}$, $\Sigma_{\parallel}$ and $\Sigma_{\perp}$ depend on median redshift of the ZH bin we are fitting to, and are listed in Table \ref{tab:cov_vals}. In order for the matter power spectrum to be used in our covariance matrix calculation, it must be converted to a galaxy power spectrum using a bias appropriate for the tracers which are outlining the dark matter. 

To estimate the galaxy bias for each of our samples, we used two separate methods. For the first fitting run only, we made use of results from the HiZels survey \citep{cochrane2017}, who perform measurements of the H${\alpha}$ emitting galaxies at bins of $z=0.8$, $1.7$, and $2.23$, binning further by the mean flux of galaxies in separate bins. We use their estimates of the bias in our covariance matrix (via the power spectrum) to perform the first set of fits. We then record the best-fit values of the biases from these fits; these values were used for the covariance matrix generation for our second fitting run. This process gave us galaxy biases of approximately $[1.47,1.45,2.12,2.10,2.13]$ for ZH bins of Z2H2, Z2H3, Z3H1, Z3H2, and Z3H3, respectively. These power spectra are then used to generate covariance matrices in multipole space, for multipoles of 00, 02, 20 (transpose) and 22:
\begin{equation}
C^{\xi}_{l_{1},l_{2}}(s_{i},s_{j}) = \frac{{\rm i}^{l_{1}+l_{2}}}{2 \pi^{2}} \int_{0}^{\infty}k^{2} \sigma_{l_{1}l_{2}}^{2}(k) \bar{j}_{l_{1}}(k s_{i})\bar{j}_{l_{2}}(k s_{j})\,{\rm d}k,
\end{equation}
where the multipole-weighted variance integral is
\begin{equation}
\sigma^{2}_{l_{1}l_{2}}(k) = \frac{(2l_{1}+1)(2l_{2}+1)}{V_{s}} \int_{-1}^{1} \left[P(k,\mu) + \frac1{\bar n}\right]^{2} L_{l_{1}}(\mu) L_{l_{2}}(\mu)\,{\rm d} \mu,
\label{eq:mult19}
\end{equation}
and the bin-averaged spherical Bessel function is
\begin{equation}
\bar{j}_{l}(k s_{i}) = \frac{4 \pi}{V_{si}} \int_{s_{i} - \Delta s /2}^{s_{i} + \Delta s /2} s^{2} j_{l}(ks)\,{\rm d}s.
\end{equation}
Here $V_{si}=4\pi[(s_i+\Delta s/2)^3-(s_i-\Delta s/2)^3]/3$,
$V_{s}$ is the volume of the entire sample, $j_{l}$ is the spherical Bessel function of the first kind, $k$ is the wavenumber, $s$ is the distance in redshift space, and $\bar{n}$ is the number density of galaxies for the sample in question.
In this case, since there is shot noise from both the data and the random catalogs, we make the replacement $1/\bar n \rightarrow 1/\bar n + 1/\bar n_{\rm R}$ in Eq.~(\ref{eq:mult19}); this increases the shot noise by a factor of $\frac43$ for $n_{\rm R}/n_{\rm D}=3$.

Our simulations are not volume-limited, and have a galaxy number density, $\bar{n}$, which is implicitly dependent on redshift, while our theoretical method to calculate the covariance matrices assumes a constant galaxy number density. To account for this, we further divided each of our ZH bins into three sub-bins by redshift. The covariance matrix of each sub-bin was calculated using the volume and number density of galaxies within that specific sub-bin. The covariance matrix of the entire ZH bin was calculated by:
\begin{equation} \label{eq:newcovariance}
C^{\xi}_{l_{1},l_{2}}(s_{i},s_{j}) = \sum_{k} w_{k}^{2} C^{\xi}_{l_{1},l_{2},k}(s_{i},s_{j}),
\end{equation}
where $k$ indicates the specific sub-bin, and:
\begin{equation}
w_{k} = \frac{v_{k}\bar{n_{k}}^{2}}{\sum_{k'}v_{k'}\bar{n_{k'}}^{2}},
\end{equation}
For our Z2 bins, the redshift cutoffs were at 0.9 and 1.1, and for Z3, they were 1.55 and 1.75. The volume of our survey area over the redshift-range of the subset $k$ is designated as $v_{k}$.

\subsection{RSD parameter fitting}
\label{subsec:rsd_parameter}

The fit on small scales follows the procedure in \cite{martens2018}, focusing on the redshift space distortion parameters. We fit on scales of 42 to 200 Mpc, with a factor of 4 lower spatial resolution than for the BAO fits; it was reduced in order to prevent small-scale fluctuations in the correlation function dominating the best-fit values. 

To calculate the theoretical fit, we use CLPT, modified on small scales by GSRSD. In order to fit a theoretically produced correlation function to our simulations, we use an extension of CLPT \citep{carlson2012}. CLPT extends perturbation theory beyond linear order to match up to quasi-linear scales of the correlation function; the Gaussian streaming model tailors the fit to behave better on small scales \citep{wang2014}. GSRSD takes as input $\sigma_{\rm FOG}^{2}$, $f_{v}$, and $b_{g}$. We treat $\sigma_{\rm FOG}^{2}$ as a fixed parameter, however we run several fits over different set values of $\sigma_{\rm FOG}^{2}$, fitting for $f_{v}$ and $b_{g}$ with each different set value. This is done to provide a more stringent test to the similarity of the observed and shuffled catalogs, since fixing $\sigma_{\rm FOG}^{2}$ shrinks the uncertainties in the fit for $f_{v}$; it should be noted that the real data collected by \WFIRST\ will be simultaneously fit for $f_{v}$, $b_{g}$, and $\sigma_{\rm FOG}^{2}$, although the scales over which these will be fit could differ from those presented here.

The observed and shuffled catalogs have a different set value of $\sigma_{\rm FOG}^{2}$ than the true catalogs, due to the spreading introduced in Eq.~(\ref{eq:obs-spread}). This additional spreading was calculated separately for each ZH bin, and was found to be 24.222, 24.324, 21.097, 21.098, and 21.080 $h^{-2}\, \mathrm{Mpc}^2$ for ZH bins of Z2H2, Z2H3, Z3H1, Z3H2, and Z3H3, respectively. We run our fits for $\sigma_{\rm FOG}^{2}$ values of 5, 20, and 35 $(h^{-1}\mathrm{Mpc})^{2}$ for the true catalogs, which is added to the additional spreading found for the observed and shuffled catalogs.

In order to produce these outputs, GSRSD also takes as input the galaxy bias, and $f_v$, which are our primary fitting parameters. The code outputs the redshift-space correlation function in terms of moments, $\xi_{0,2,4}(r)$, which is directly comparable to our simulated correlation functions. 

We have repeated the RSD parameters for alternative scale ranges (e.g., 60--180 Mpc) and found that the true/observed/shuffled parameter shifts change by $<0.1\%$.


\subsection{BAO Parameters}
\label{subsec:BAO}

\WFIRST\ will perform accurate BAO measurements up to $z \sim 1.9$ using H$\alpha$, and to higher redshifts using [O{\,\sc iii}] and [O{\,\sc ii}] emitters, pinning down the expansion rate of the Universe, $H(z)$ and the angular diameter distance, $D_{A}(z)$. However, the line blending effect studied in this paper will potentially bias the results. To quantify this bias and track down how much arises from the redshift error PDF versus its correlation with large-scale structure, we will compare the BAO parameters measured from our true, observed, and shuffled catalogs. In this subsection, we first introduce the BAO model, and then discuss about our fitting process. They do {\em not} include reconstruction, which we leave to future analysis.

The BAO fits are performed over scales from 60 to 200 Mpc, and are intended to calculate the expansion rate of the Universe at a specific redshift, $H(z)$, and the angular diameter distance to that redshift, $D_{A}(z)$. We follow the standard BAO convention in defining
\begin{equation}
\alpha_{\parallel} = \frac{(H(z)r_{d})^{\mathrm{fid}}}{H(z)r_{d}} ~~~{\rm and}~~~ \alpha_{\perp} = \frac{D_{A}(z)r_{d}^{\mathrm{fid}}}{D^{\mathrm{fid}}_{A}(z)r_{d}},
\end{equation}
where here, a superscript of ``$\mathrm{fid}$'' indicates a fiducial value, and $r_{d}$ is the sound horizon at the kinetic decoupling epoch. Given $P(k,\mu)$, we calculate the multipole moments
\begin{equation}
P_{l}(k) = \frac{2l+1}{2}\int_{-1}^1 P(k,\mu) L_l(\mu)d\mu~,
\end{equation}
and then transform them into real space as
\begin{equation}
\xi_l(s) = \frac{i^l}{2\pi^2}\int \frac{dk}{k}\,k^3 P_l(k)j_l(ks)~,
\label{eq:xils}
\end{equation}
so that the correlation function is expressed as
\begin{equation}
\xi(s,\mu) = \sum_{l}\xi_l(s)L_l(\mu),
\end{equation}
where we only sum to $l=4$.

We fit to the same BAO model in \cite{ross2016},
\begin{align}
\xi_{0,{\rm mod}}(s) =& B_0\xi_{\mu 0}(s,\alpha_\perp,\alpha_\parallel)+\frac{a_{01}}{s^2}+\frac{a_{02}}{s}+a_{03}~~~{\rm and}\\
\xi_{2,{\rm mod}}(s) =& \frac{5}{2}[B_2\xi_{\mu 2}(s,\alpha_\perp,\alpha_\parallel)-B_0\xi_{\mu 0}(s,\alpha_\perp,\alpha_\parallel)]\nonumber \\
&+\frac{a_{21}}{s^2}+\frac{a_{22}}{s}+a_{23},
\end{align}
where $\xi_{\mu 0},\xi_{\mu 0}$ are $\mu$-averaged $\xi(s,\mu)$, defined by
\begin{equation}
\xi_{\mu 0}(s,\alpha_\perp,\alpha_\parallel)=\int_0^1 d\mu\,\xi(s',\mu')
\end{equation}
and
\begin{equation}
\xi_{\mu 2}(s,\alpha_\perp,\alpha_\parallel)=\int_0^1 d\mu\,3\mu'^2 \xi(s',\mu'),
\end{equation}
with $\mu'=\mu\alpha_\parallel/\sqrt{\mu^2\alpha_\parallel^2+(1-\mu^2)\alpha_\perp^2}$, and $s'=s\sqrt{\mu^2\alpha_\parallel^2+(1-\mu^2)\alpha_\perp^2}$. Parameters $B_i$ and $a_{ij}$ set the size of the BAO and the broadband feature.

We fit the model by minimizing the $\chi^2$ in Eq.~(\ref{eq:chi_squared}). Since the model depends linearly on the parameters $B_i$ and $a_{ij}$, the $\chi^2$ is a quadratic function of them. For each given pair of $(\alpha_\perp,\alpha_\parallel)$, one can calculate the other 8 parameters where the $\chi^2$ takes a minimum. Let $\bm{p}^{(8)}$ be the vector of the 8 parameters, and $p_i$ be its $i$-th component; then the $\chi^2$ can be written as
\begin{equation}
\chi^2(\alpha^\perp,\alpha^\parallel;\bm{p}^{(8)}) = \bm{d}^{\rm T}{\mathbfss C}^{-1}\bm{d} -2\sum_{i=1}^8 p_i J_i + \sum_{i,j=1}^8p_i p_j K_{ij}.
\end{equation}
Here $\bm{d}$ is the data vector, and we define
$J_i=\bm{m}_i^{\rm T}{\mathbfss C}^{-1}\bm{d}$ and
$K_{ij}=\bm{m}_i^{\rm T}{\mathbfss C}^{-1}\bm{m}_j$,
where $\bm{m}_i$ is the model vector if the parameter $p_i$ is set to be 1 and all other seven components of $\bm{p}^{(8)}$ are set to be 0. The minimum of $\chi^2$ is obtained if
$p_i^{(8)} = [{\mathbfss K}^{-1}]_{ij}J_j$.
We first use Eq.~(\ref{eq:xils}) to generate the theoretical $\xi_{0,2,4}(s)$ with 5000 $s$-values logarithmically spaced in [1,240] Mpc for each ZH bin. Several techniques have been applied to improve the accuracy of the integration. Firstly, we use 50,000 $k$-values logarithmically sampled from $10^{-4}$ to 100$\,$Mpc$^{-1}$. Secondly, a window function that has continuous first and second derivatives\footnote{See Eq.~(C.1) in \cite{2016JCAP...09..015M}. Here we have $k_{\rm max}=100\,$Mpc$^{-1}$ and the window applies from 80 to 100$\,$Mpc$^{-1}$.} is applied to the high-$k$ end of the $k$-dependent function $k^3P_l(k)$, to remove the high-frequency ringing appearing in $\xi_l(s)$. Finally, we multiply $j_l(ks)$ by a factor of $\sin(ks\Delta\ln k/2)/(ks\Delta\ln k/2)$ to effectively average out the contribution from the rapidly oscillating spherical Bessel functions at large $k$. During each iteration of the minimization, $\xi_l(s')$ is calculated from the pre-calculated theoretical $\xi_l(s)$ using the cubic interpolation method. The minimization uses the \texttt{scipy.optimize.minimize} routine \citep{scipy} with the Nelder-Mead method, and the initial guess of $(\alpha_\perp,\alpha_\parallel)$ is always $(1,1)$.

We have tried adjusting the BAO fitting range, varying the lower cutoff from the default (60 Mpc) to 50, 70, or 80 Mpc; the BAO scale parameters change by $<0.1\%$ in all cases.

\subsection{Fitting Results}
Here, we show the resulting parameters that were fit to each correlation function, and discuss trends in the results. For each ZH bin, we have 6 separate regions, which serve as 6 realizations of simulation data. We average the correlation functions from all sectors: 
\begin{equation}
\xi_{\mathrm{avg}} = \frac{\xi_{1}+\xi_{2}+\xi_{3}+\xi_{4}+\xi_{5}+\xi_{6}}{6}.
\end{equation}
We then fit parameters to $\xi_{\mathrm{avg}}$, which are denoted as $\bar{p}$ for a given parameter $p$. These fits provide the listed parameters in Tables \ref{tab:finalparams_RSD} and \ref{tab:finalparams_BAO}, as well as the listed $\chi^{2}$ per degree of freedom.

We find the error bars for each parameter using the jack-knife method, where we construct 6 separate combinations of the correlation functions, with each combination containing all regions except for one:
\begin{equation}
\bar{\xi}_{i} = \frac{1}{5}\sum_{j=1,j\neq i}^{6} \xi_{i}.
\end{equation}
We then separately fit the correlation functions $\bar{\xi}_{i}$ for the BAO and RSD parameters. We show error bars which are derived from the variance of the set of parameters of our combinations:
\begin{equation} \label{eq:jack_std}
\sigma_{\bar{p}} = \sqrt{\frac{5}{6} \sum_{i=1}^{6} (\bar{p}_{i} - \bar{p})^{2}}.
\end{equation}
The jack-knife method, while providing an estimate of the errors in our parameters due to deviations in the correlation functions, is limited by our small sample set of only 6 realizations. Table \ref{tab:finalparams_RSD} displays the $f_{v}$ and $b_{g}$ best-fit parameters for each ZH bin, and each of the true, observed, and shuffled catalogs. There are three columns in the table, each indicating the best-fits in cases of varying values for the Finger of God spreading, $\sigma_{\rm FOG}^{2}$. We list the true catalog value of $\sigma_{\rm FOG}^{2}$ for the True catalog in Table~\ref{tab:finalparams_RSD}. These values were chosen to represent a wide range of values for $\sigma_{\rm FOG}^{2}$ for our samples, and to show how their choice affects the differences between the observed and shuffled catalogs. The error bars here are calculated using Eq.~(\ref{eq:jack_std}). Table \ref{tab:finalparams_RSD} also shows, for each ZH bin, the percent error between the fitted parameters between the observed and shuffled catalogs. This is calculated by:
\begin{equation} \label{eq:deltap}
\Delta p \equiv \frac{p_{\mathrm{obs}} - p_{\mathrm{shf}}}{0.5(p_{\mathrm{obs}}+p_{\mathrm{shf}})}\times 100\%,
\end{equation}
where $p$ is either $b_{g}$, $f_{v}$, $\alpha_{\perp}$ or $\alpha_{\parallel}$.

\setlength{\tabcolsep}{0.1em}
\begin{sidewaystable*}\centering
\ra{1.3}
\setlength{\tabcolsep}{3.5pt}
\caption{Here we display the best fit parameters for the growth of structure parameter $f_{v}$ and the galaxy bias $b_{g}$, for each ZH bin and redshift catalog. We calculate $\%$ error (Eq.~\ref{eq:deltap}) to indicate the agreement between the shuffled and observed parameters. The uncertainties provided by the jack-knife sampling have a fractional error of $\sqrt{2/5}$, due to the sample size of 6 regions.}
\label{tab:finalparams_RSD}
\begin{tabular}{@{}cccccccccccc@{}}\toprule
& \multicolumn{3}{c}{\textbf{$\sigma_{\rm FOG}^{2} = 5$ $(h^{-1}\mathrm{Mpc})^{2}$}} & \phantom{abc} & \multicolumn{3}{c}{\textbf{$\sigma_{\rm FOG}^{2} = 20$ $(h^{-1}\mathrm{Mpc})^{2}$}} & \phantom{abc} & \multicolumn{3}{c}{\textbf{$\sigma_{\rm FOG}^{2} = 35$ $(h^{-1}\mathrm{Mpc})^{2}$}}\\
\cmidrule{2-4} \cmidrule{6-8} \cmidrule{10-12}
& $b_{g}$ & $f_{v}$ & $\chi^{2}$/d.o.f && $b_{g}$ & $f_{v}$ & $\chi^{2}$/d.o.f && $b_{g}$ & $f_{v}$ & $\chi^{2}$/d.o.f \\ \midrule
\textbf{Z2H2}\\
True & 1.4817 $\pm$ 0.0139 & 0.7080 $\pm$ 0.0148 & 1.219 && 1.4658 $\pm$ 0.0139 & 0.7593 $\pm$ 0.0147 & 1.285 && 1.4487 $\pm$ 0.0137 & 0.8129 $\pm$ 0.0149 & 1.819 \\
Observed & 1.4938 $\pm$ 0.0120 & 0.6878 $\pm$ 0.0126 & 1.674 && 1.4729 $\pm$ 0.0119 & 0.7489 $\pm$ 0.0127 & 1.359 && 1.4497 $\pm$ 0.0119 & 0.8135 $\pm$ 0.0128 & 1.813 \\
Shuffled & 1.4934 $\pm$ 0.0120 & 0.6898 $\pm$ 0.0113 & 1.618 && 1.4724 $\pm$ 0.0118 & 0.7504 $\pm$ 0.0115 & 1.336 && 1.4493 $\pm$ 0.0119 & 0.8154 $\pm$ 0.0115 & 1.828 \\
$\% $Error & \textbf{0.03 $\pm$ 0.10 $\%$} & \textbf{-0.28 $\pm$ 0.32 $\%$} & - && \textbf{0.04 $\pm$ 0.12 $\%$} & \textbf{-0.20 $\pm$ 0.30 $\%$} & - && \textbf{0.03 $\pm$ 0.11 $\%$} & \textbf{-0.22 $\pm$ 0.25 $\%$} & - \\
\textbf{Z2H3}\\
True & 1.4585 $\pm$ 0.0139 & 0.6891 $\pm$ 0.0200 & 1.512 && 1.4429 $\pm$ 0.0138 & 0.7380 $\pm$ 0.0199 & 1.574 && 1.4262 $\pm$ 0.0139 & 0.7892 $\pm$ 0.0201 & 1.971 \\
Observed & 1.4643 $\pm$ 0.0107 & 0.6734 $\pm$ 0.0161 & 1.464 && 1.4444 $\pm$ 0.0107 & 0.7298 $\pm$ 0.0161 & 1.376 && 1.4227 $\pm$ 0.0107 & 0.7898 $\pm$ 0.0164 & 1.803 \\
Shuffled & 1.4676 $\pm$ 0.0127 & 0.6707 $\pm$ 0.0178 & 1.617 && 1.4479 $\pm$ 0.0125 & 0.7271 $\pm$ 0.0179 & 1.478 && 1.4262 $\pm$ 0.0125 & 0.7871 $\pm$ 0.0180 & 1.847 \\
$\% $Error & \textbf{-0.22 $\pm$ 0.37 $\%$} & \textbf{0.40 $\pm$ 0.51 $\%$} & - && \textbf{-0.25 $\pm$ 0.39 $\%$} & \textbf{0.36 $\pm$ 0.49 $\%$} & - && \textbf{-0.24 $\pm$ 0.38 $\%$} & \textbf{0.34 $\pm$ 0.41 $\%$} & - \\
\textbf{Z3H1}\\
True & 2.1512 $\pm$ 0.0072 & 0.7631 $\pm$ 0.0065 & 2.373 && 2.1294 $\pm$ 0.0074 & 0.8287 $\pm$ 0.0066 & 2.112 && 2.1056 $\pm$ 0.0073 & 0.8974 $\pm$ 0.0066 & 2.949 \\
Observed & 2.1531 $\pm$ 0.0054 & 0.7561 $\pm$ 0.0076 & 2.525 && 2.1256 $\pm$ 0.0052 & 0.8305 $\pm$ 0.0078 & 1.956 && 2.0956 $\pm$ 0.0054 & 0.9090 $\pm$ 0.0079 & 2.992 \\
Shuffled & 2.1539 $\pm$ 0.0060 & 0.7574 $\pm$ 0.0064 & 2.362 && 2.1268 $\pm$ 0.0059 & 0.8317 $\pm$ 0.0064 & 1.952 && 2.0967 $\pm$ 0.0058 & 0.9104 $\pm$ 0.0065 & 3.156 \\
$\% $Error & \textbf{-0.04 $\pm$ 0.09 $\%$} & \textbf{-0.18 $\pm$ 0.41 $\%$} & - && \textbf{-0.06 $\pm$ 0.09 $\%$} & \textbf{-0.14 $\pm$ 0.41 $\%$} & - && \textbf{-0.05 $\pm$ 0.07 $\%$} & \textbf{-0.15 $\pm$ 0.36 $\%$} & - \\
\textbf{Z3H2}\\
True & 2.1307 $\pm$ 0.0066 & 0.7476 $\pm$ 0.0070 & 2.149 && 2.1095 $\pm$ 0.0066 & 0.8093 $\pm$ 0.0073 & 1.939 && 2.0867 $\pm$ 0.0063 & 0.8740 $\pm$ 0.0073 & 2.593 \\
Observed & 2.1335 $\pm$ 0.0045 & 0.7483 $\pm$ 0.0087 & 2.095 && 2.1079 $\pm$ 0.0045 & 0.8170 $\pm$ 0.0088 & 1.939 && 2.0798 $\pm$ 0.0043 & 0.8894 $\pm$ 0.0092 & 2.980 \\
Shuffled & 2.1350 $\pm$ 0.0039 & 0.7467 $\pm$ 0.0097 & 2.211 && 2.1095 $\pm$ 0.0039 & 0.8156 $\pm$ 0.0100 & 2.069 && 2.0812 $\pm$ 0.0038 & 0.8881 $\pm$ 0.0100 & 3.121 \\
$\% $Error & \textbf{-0.07 $\pm$ 0.10 $\%$} & \textbf{0.21 $\pm$ 0.34 $\%$} & - && \textbf{-0.08 $\pm$ 0.11 $\%$} & \textbf{0.17 $\pm$ 0.33 $\%$} & - && \textbf{-0.07 $\pm$ 0.10 $\%$} & \textbf{0.15 $\pm$ 0.29 $\%$} & - \\
\textbf{Z3H3}\\
True & 2.1535 $\pm$ 0.0121 & 0.7217 $\pm$ 0.0112 & 0.946 && 2.1332 $\pm$ 0.0124 & 0.7737 $\pm$ 0.0112 & 1.260 && 2.1112 $\pm$ 0.0123 & 0.8276 $\pm$ 0.0114 & 1.945 \\
Observed & 2.1650 $\pm$ 0.0105 & 0.7152 $\pm$ 0.0116 & 1.425 && 2.1420 $\pm$ 0.0104 & 0.7710 $\pm$ 0.0117 & 1.762 && 2.1171 $\pm$ 0.0103 & 0.8290 $\pm$ 0.0120 & 2.543 \\
Shuffled & 2.1616 $\pm$ 0.0126 & 0.7079 $\pm$ 0.0109 & 1.365 && 2.1390 $\pm$ 0.0126 & 0.7632 $\pm$ 0.0111 & 1.522 && 2.1143 $\pm$ 0.0127 & 0.8214 $\pm$ 0.0114 & 2.118 \\
$\% $Error & \textbf{0.15 $\pm$ 0.30 $\%$} & \textbf{1.03 $\pm$ 1.21 $\%$} & - && \textbf{0.14 $\pm$ 0.30 $\%$} & \textbf{1.01 $\pm$ 1.11 $\%$} & - && \textbf{0.13 $\pm$ 0.31 $\%$} & \textbf{0.92 $\pm$ 1.09 $\%$} & - \\
\bottomrule
\end{tabular}
\end{sidewaystable*}

\begin{figure}
\centering
\includegraphics[width=\columnwidth,height=!,keepaspectratio]{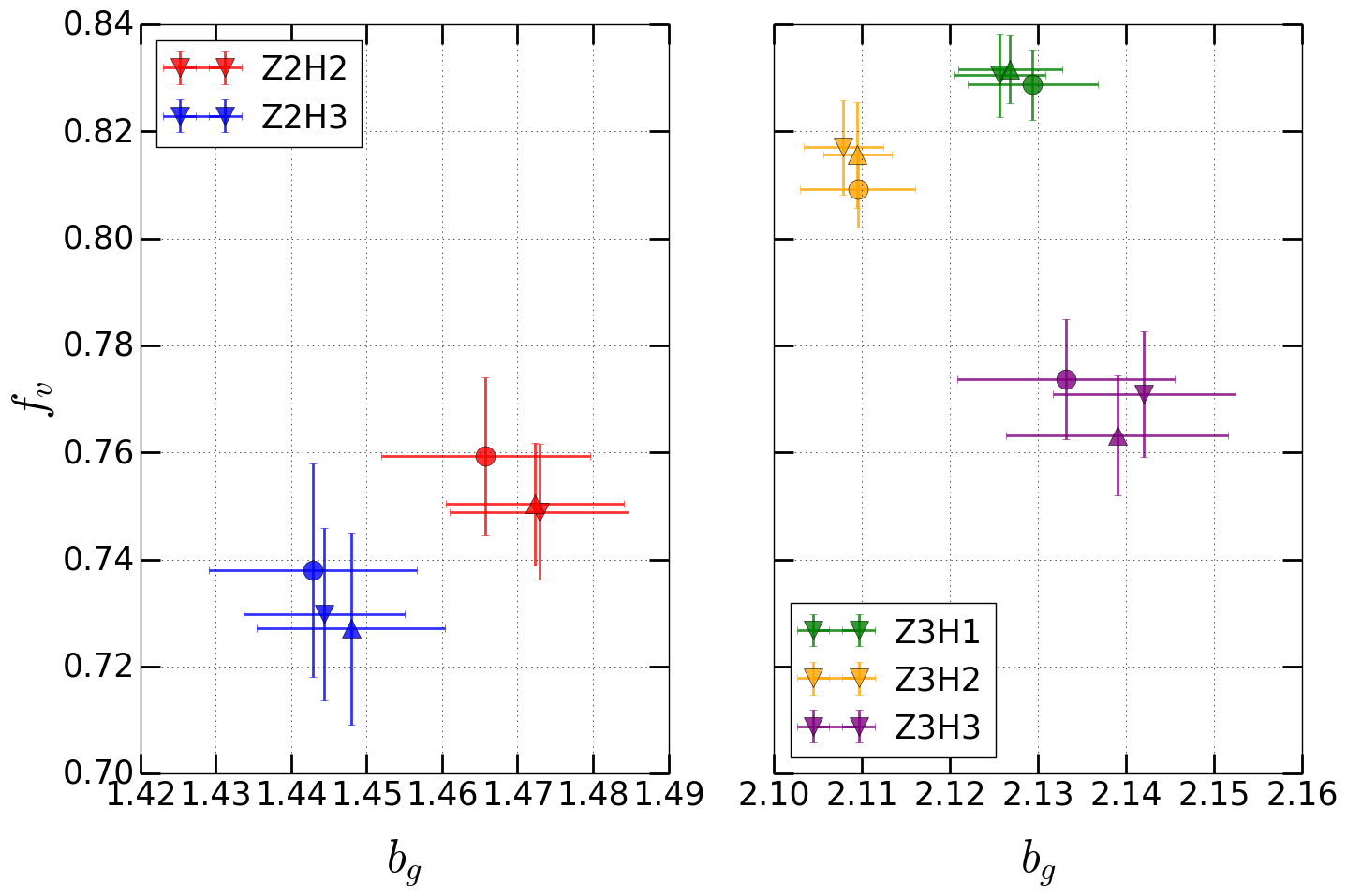}
\caption{For each ZH bin and Observed, True, and Shuffled catalog, we plot the fitted values for $f_{v}$ and $b_{g}$. Circles indicate the True catalog, V's indicate the Observed catalog, and triangles indicate the Shuffled catalog. For this fitting calculation, $\sigma_{\rm FOG}^{2}$ was set to $20 \left[ h^{-1}\mathrm{Mpc}\right]^{2}$ for the True catalog.}
\label{fig:param1_15}
\end{figure}

In Figure \ref{fig:param1_15}, we plot the best-fit parameters for a visual comparison of the different RSD results, we show the full RSD best-fit parameters in Table \ref{tab:finalparams_RSD}, and the best-fit BAO parameters are displayed in Table \ref{tab:finalparams_BAO}. 


Systematic errors in the mean parameter fits could arise from a few assumptions made in the modeling. Mean redshifts were calculated for each ZH bin, and were used to determine the parameters $\Sigma_\perp,\Sigma_\parallel,\Sigma_s$, and generate the matter power spectra. Since our samples are not volume-limited, the redshift dependence of the galaxy number densities in each bin may introduce errors in the modeling.

Some of the large $\chi^2/$d.o.f.\ values result from the combination of the bin-to-bin scatter in $\xi_l(s)$ and the strong correlations between neighboring radial bins (\ie the off-diagonal terms of the covariance matrix). We suspect that the analytic covariance matrix may not be adequate for describing these rapidly oscillating modes in $\xi_l(s)$. Although we attempt to generalize it to a non-volume-limited sample, we still must estimate a galaxy bias for the input power spectrum, and the rapidly oscillating modes correspond to high $k$ where non-linear biasing may be important. The covariance matrix uses a Gaussian approximation, neglecting the galaxy trispectrum and super-sample variance effects, which may also break down for these modes. In any case, these highly oscillating modes are orthogonal to the broadband modes that dominate $f_v$, and changes in our treatment of the covariance matrix in the preparation of this paper had a small impact on changes in the parameter shifts.



\begin{table*}\centering
\ra{1.3}
\setlength{\tabcolsep}{12pt}
\caption{The best-fit parameters for the BAO shift parameters $\alpha_{\parallel}$ and $\alpha_{\perp}$, and their $\chi^{2}$ values. We calculate $\%$ error (Eq. \ref{eq:deltap}) to indicate the agreement between the shuffled and observed parameters. The uncertainties provided by the jack-knife sampling have a fractional error of $\sqrt{2/5}$, due to the sample size of 6 regions.}
\label{tab:finalparams_BAO}
\begin{tabular}{@{}ccccccccc@{}}\toprule
& \multicolumn{3}{c}{\textbf{BAO Parameters}} \\
& $\alpha_{\parallel}$ & $\alpha_{\perp}$ & $\chi^{2}$/d.o.f \\ \midrule
\textbf{Z2H2}\\
True & 0.9858 $\pm$ 0.0111 & 0.9800 $\pm$ 0.0066 & 0.484 \\
Observed & 0.9920 $\pm$ 0.0092 & 0.9780 $\pm$ 0.0062 & 0.678 \\
Shuffled & 0.9895 $\pm$ 0.0098 & 0.9792 $\pm$ 0.0054 & 0.807 \\
$\% $Error & \textbf{0.26 $\pm$ 0.17 $\%$} & \textbf{-0.12 $\pm$ 0.11 $\%$} & - \\
\textbf{Z2H3}\\
True & 1.0045 $\pm$ 0.0166 & 0.9739 $\pm$ 0.0088 & 0.486 \\
Observed & 0.9977 $\pm$ 0.0154 & 0.9708 $\pm$ 0.0076 & 0.826 \\
Shuffled & 0.9986 $\pm$ 0.0153 & 0.9726 $\pm$ 0.0078 & 0.871 \\
$\% $Error & \textbf{-0.08 $\pm$ 0.24 $\%$} & \textbf{-0.19 $\pm$ 0.11 $\%$} & - \\
\textbf{Z3H1}\\
True & 0.9614 $\pm$ 0.0048 & 0.9763 $\pm$ 0.0038 & 1.537 \\
Observed & 0.9598 $\pm$ 0.0044 & 0.9799 $\pm$ 0.0039 & 1.362 \\
Shuffled & 0.9586 $\pm$ 0.0051 & 0.9793 $\pm$ 0.0041 & 1.241 \\
$\% $Error & \textbf{0.13 $\pm$ 0.19 $\%$} & \textbf{0.07 $\pm$ 0.05 $\%$} & - \\
\textbf{Z3H2}\\
True & 0.9592 $\pm$ 0.0025 & 0.9753 $\pm$ 0.0041 & 1.668 \\
Observed & 0.9668 $\pm$ 0.0042 & 0.9690 $\pm$ 0.0035 & 1.427 \\
Shuffled & 0.9638 $\pm$ 0.0025 & 0.9699 $\pm$ 0.0037 & 1.468 \\
$\% $Error & \textbf{0.31 $\pm$ 0.23 $\%$} & \textbf{-0.10 $\pm$ 0.10 $\%$} & - \\
\textbf{Z3H3}\\
True & 0.9625 $\pm$ 0.0110 & 0.9757 $\pm$ 0.0033 & 1.286 \\
Observed & 0.9488 $\pm$ 0.0088 & 0.9702 $\pm$ 0.0064 & 1.109 \\
Shuffled & 0.9504 $\pm$ 0.0080 & 0.9721 $\pm$ 0.0060 & 1.189 \\
$\% $Error & \textbf{-0.17 $\pm$ 0.55 $\%$} & \textbf{-0.19 $\pm$ 0.15 $\%$} & - \\
\bottomrule
\end{tabular}
\end{table*}

\section{Measuring and Correcting Error Terms}
\label{sec:mitigation}

As discussed in Sec. \ref{sec:objectives}, \WFIRST\ has a requirement to meet, and preferably exceed, the observational systematic error requirements in the SRD for measured parameters; specifically, the parameters measured in this work are the BAO shift parameters $\alpha_{\parallel}$ and $\alpha_{\perp}$, and the growth of structure parameter $f_{v}$. The error limits are 0.41\%\ for $\alpha_\perp$, 0.74\%\ for $\alpha_\parallel$, and 0.74\%\ for $f_v$ (there is no top-level requirement on the systematic errors to the galaxy bias $b_{g}$, as we will marginalize over it in estimating cosmological parameters). In this section, we discuss the breakdown of line blending errors within the structure of our catalogs, compare our results to the error budget of the SRD, and briefly introduce potential mitigation techniques that could limit the negative effects of line blending on the observed redshift. Note that in our discussions of ``percent of error budget'' used, we will use root-sum-square (RSS) error budgeting, so that an effect whose amplitude is 50\%\ of the maximum allowed error is considered to use 25\%\ of the total budget.

In this paper, we have presented the fit parameters for three separate catalogs of redshift: true, observed, and shuffled. Their derivation is described in detail in Sec. \ref{subsec:binning}. The purpose of these catalogs is to separate the redshift errors introduced by line blending into two distinct sources. The difference between the parameters of the observed and true catalogs (which we wish to eventually mitigate) can be written as: 
\begin{equation} \label{eq:miti_one}
p_{\mathrm{obs}} - p_{\mathrm{true}} = (p_{\mathrm{obs}} - p_{\mathrm{shuffled}}) + (p_{\mathrm{shuffled}} - p_{\mathrm{true}}).
\end{equation}
The first part of Eq.~(\ref{eq:miti_one}) is comprised of the difference between the parameters of the observed and shuffled catalogs. By construction, these two catalogs have identical {\em distributions} of redshift errors; however, in the shuffled catalog, the line-blending redshift difference has been uncoupled from the specific galaxy that generated it, destroying the correlation between [N\,{\sc ii}]/H$\alpha$ and galaxy environment. Therefore, this term describes the effect of this correlation. It is this term that -- although small -- is not amenable to mitigation by measuring the one-point PDF of galaxy properties.

In contrast, the second part of Eq.~(\ref{eq:miti_one}) depends only on the one-point properties of galaxies, such as their metallicity distribution. Because there is {\em no} large-scale structure information encoded in this term, it is more straightforward to mitigate via detailed observations of a small number of galaxies. We will analyze the error encompassed by these terms separately, in order to discuss the necessary mitigation techniques. 

\subsection{Effect of correlations between [N\,{\sc ii}]/H$\alpha$ and large-scale environment}

\begin{figure}
\centering
\includegraphics[width=\columnwidth,height=!,keepaspectratio]{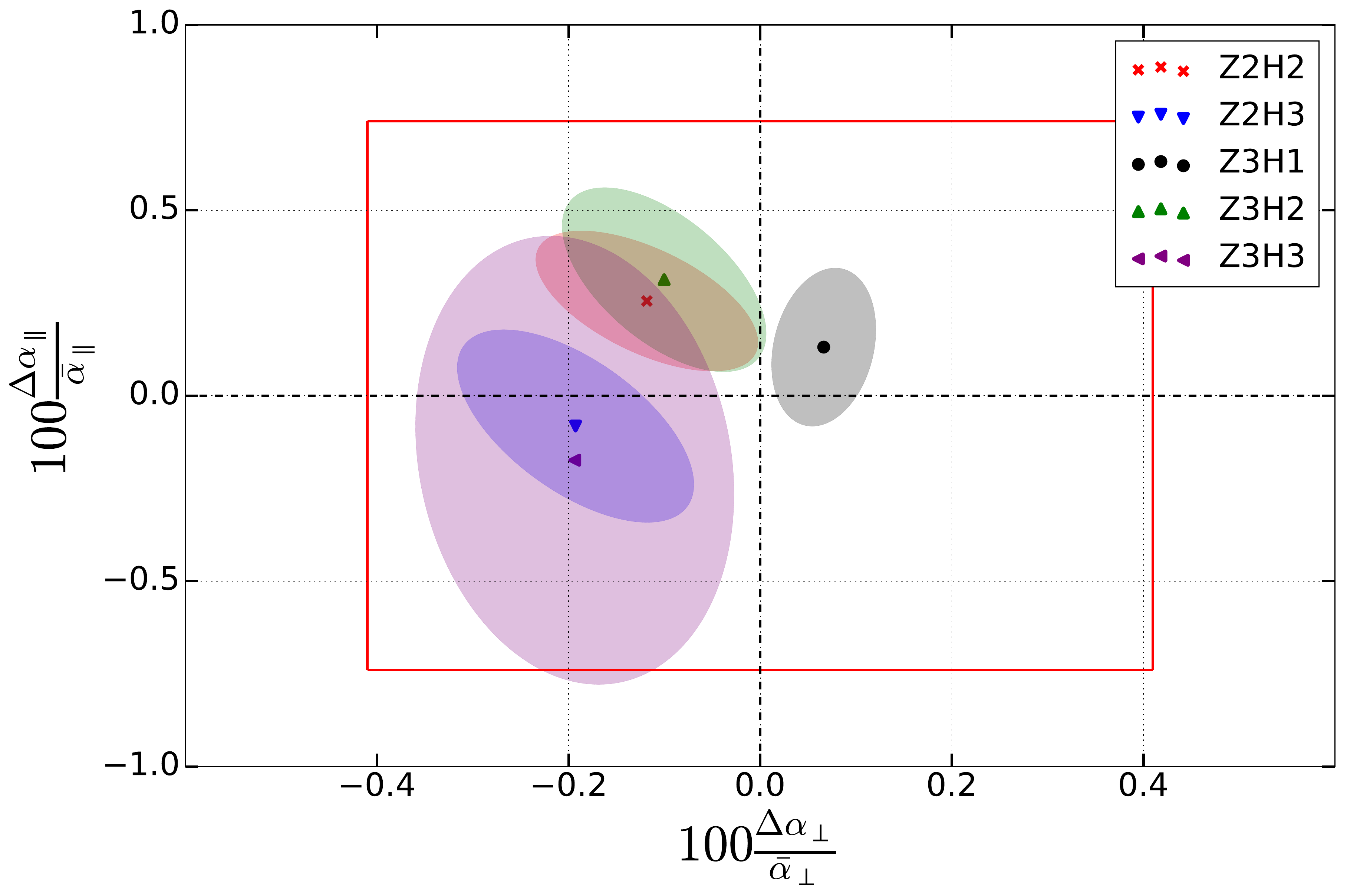}
\caption{For each ZH bin, we show the percent difference of $\alpha_{\perp}$ and $\alpha_{\parallel}$ relative to the systematic error budget (red box) of \WFIRST\ for each parameter. The contours show the spread of fits for all 6 jack-knife combinations for that specific ZH bin, as referenced in Eq. \ref{eq:jack_std}, while the central values are calculated from the fits of the average of all regions.}
\label{fig:diff_plot_BAO}
\end{figure}

We first analyze the difference between the shuffled and observed catalogs. Although we have fit a variety of redshift and flux bins, we will confine our numerical analysis here to the Z2H2 and Z3H2 bins, since the limiting flux of $1.3 \times 10^{-16}$ erg cm$^{-2}$ s$^{-1}$ is closest to the planned \WFIRST\ flux limit.\footnote{This flux limit varies somewhat as a function of wavelength and ecliptic latitude, and is subject to change during future optimization.} For the BAO parameters $\alpha_\perp$ and $\alpha_\parallel$, the best-fit values are listed in Table \ref{tab:finalparams_BAO}, along with the percentage difference between the observed and shuffled best-fit values. These differences are also shown graphically in Fig. \ref{fig:diff_plot_BAO}. The magnitude of the differences in $\alpha_\parallel$ fall generally between $0.1-0.3\%$, while for $\alpha_\perp$ they range from $0.1-0.2\%$. The errors on these differences are provided by the jack-knife method for all 6 regions we fit for. It is important to note that the jack-knife errors are based on a small sample size of 6 regions (for which the expected fluctuations in error bars are $\pm1/\sqrt{2(6-1)}\approx 32\%$); this could explain why in some cases we have a larger error bar for a ZH bin with a smaller sample size.


Specifically, we find that $\Delta\alpha_{\parallel} = 0.31 \pm 0.23 \%$ ($0.26 \pm 0.17 \%$) for Z3H2 (Z2H2), and $\Delta\alpha_{\perp} = -0.10 \pm 0.10 \%$ ($-0.12 \pm 0.11 \%$) for Z3H2 (Z2H2). Using the largest value of each, this corresponds to $\Delta\alpha_{\parallel}$ using $18\%$ of the systematic error budget, while $\Delta\alpha_{\perp}$ uses $9\%$. This is a small percentage of the systematic error budget, especially considering that it is for the unmitigated result. We can also compare these errors to a previous estimate of the line blending effect: \citet{faisst2017} looked at the overall redshift errors due to line blending, and propagated it the calculation of the BAO parameters; they predicted an upper estimate of these errors to be in the range of $0.5-1.6\%$ for the $\alpha_{\parallel}$ and $\alpha_{\perp}$ parameters. Our clustering analysis has shown that this estimate is indeed an upper limit, as the values we have found due to clustering are significantly lower.

The error from this portion of Eq.~(\ref{eq:miti_one}) can be potentially reduced if the [N{\,\sc ii}]/H$\alpha$ ratio can be predicted from other available data. For example, the broadband LSST+\WFIRST\ photometry (which extends into the rest-frame optical) allows estimates of the stellar mass $M_\star$, and observations of a small fraction of the \WFIRST\ sources with high-resolution ground-based NIR spectrographs would enable the correlation between [N{\,\sc ii}]/H$\alpha$ ratio to be determined as a function of redshift $z$ and inferred stellar mass $M_\star$. After applying this correction, the remaining systematic error from line blending would be associated only with the residuals from the [N{\,\sc ii}]/H$\alpha$ vs.\ $(z,M_\star)$ fit and their correlation with large-scale environment. Future work will be required to determine how much the systematic errors can be mitigated by this method.

\begin{figure}
\centering
\includegraphics[width=\columnwidth,height=!,keepaspectratio]{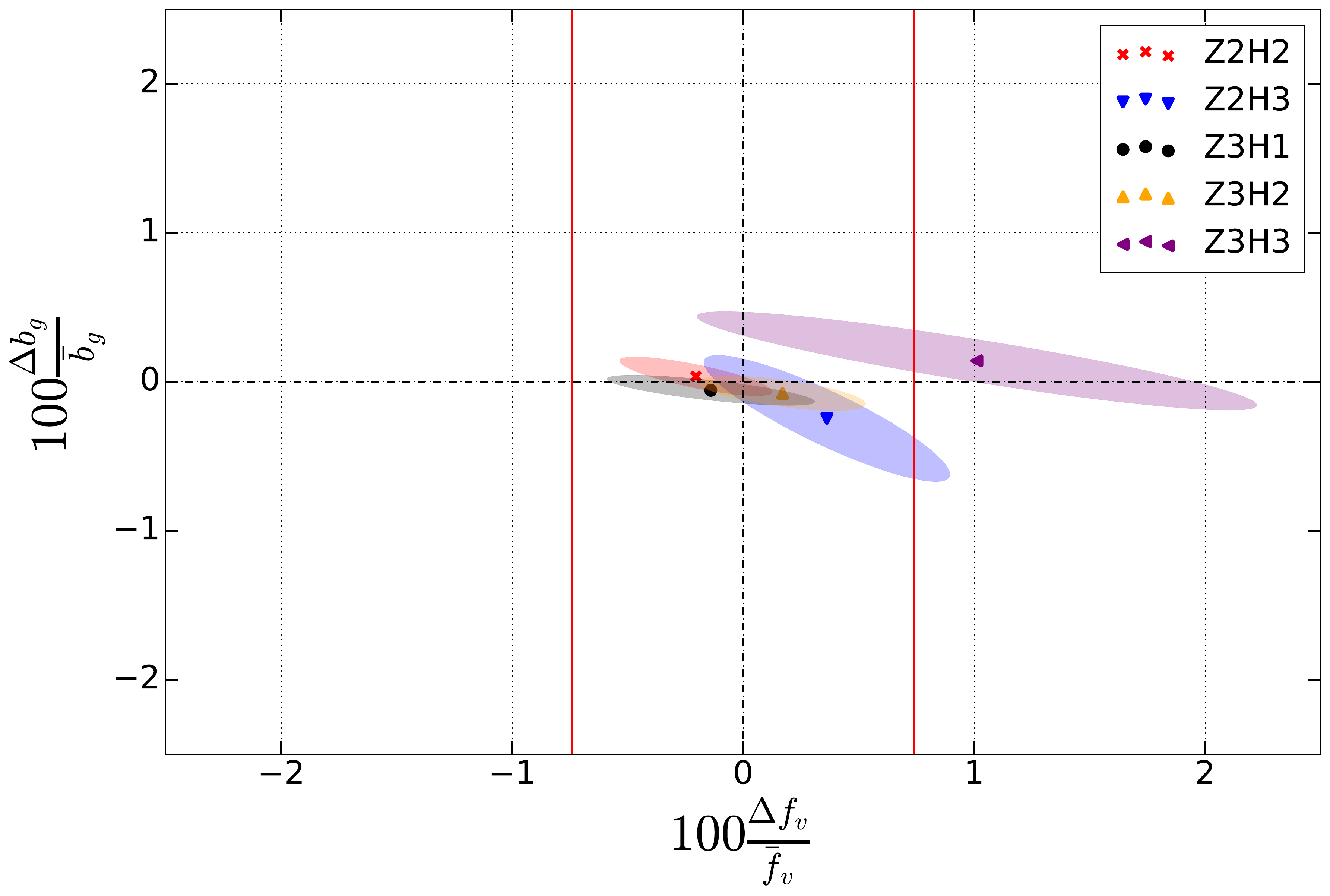}
\caption{For each ZH bin, we show the percent difference of $f_{v}$ and $b_{g}$ relative to the systematic error budget of \WFIRST\ for $f_{v}$, represented by the solid red lines. For this fitting calculation, $\sigma_{\rm FOG}^{2}$ was set to $20 \left[ h^{-1}\mathrm{Mpc}\right]^{2}$ for the True catalog. The contours show the spread of fits for all 6 jack-knife combinations for that specific ZH bin, as referenced in Eq. \ref{eq:jack_std}, while the central values are calculated from the fits of the average of all regions.}
\label{fig:diff_plot_RSD}
\end{figure}

We display the best-fit parameters for the RSD fits in Table \ref{tab:finalparams_RSD}, with the percentage difference between the observed and shuffled best-fit values also displayed in Fig. \ref{fig:diff_plot_RSD}. We have fit $f_{v}$ given several different values of $\sigma_{\rm FOG}^{2}$, and found consistency in the percentage errors in each case; here we use the values from the fit for the $\sigma_{\rm FOG}^{2} = 20 h^{-2}\mathrm{Mpc}^{2}$ bin. The differences in $f_{v}$ range between $0.14-0.36\%$, with the exception of the Z3H3 bin, which has an error of $1.01 \pm 1.11\%$ (the error bar is large due to the high flux limit and corresponding small sample size). For purposes of comparing to the SRD, we find that the error for $f_{v}$ is $0.17 \pm 0.33 \%$ ($-0.20 \pm 0.30\%$) for Z3H2 (Z2H2). For Z2H2, the larger difference, this corresponds to $f_{v}$ using $7\%$ of the systematic error budget, which again is a small enough offset that mitigations may not be required. If necessary, the aforementioned mitigations based on measurement of the [N{\,\sc ii}]/H$\alpha$ vs.\ $(z,M_\star)$ relation could reduce it even further.

\subsection{Effects of the one-point redshift error PDF}

Next, we evaluate the differences between the true and shuffled catalogs, which constitutes the second term in Eq.~(\ref{eq:miti_one}). We find that the difference for $\alpha_{\parallel}$ is approximately $-0.48\%$ ($-0.37\%$) for the Z3H2 (Z2H2) bin, which constitutes 42$\%$ of the systematic error (for the largest case). This error is similar for $\alpha_{\perp}$ at $0.56\%$ ($0.08\%$) for Z3H2 (Z2H2), but the lower error budget means it constitutes $187\%$ of the total allotment (again for the largest case). In order to move these errors below $25\%$ of the SRD limits, this would require a reduction in the errors by factors of approximately 1.3 and 2.8 for $\alpha_{\parallel}$ and $\alpha_{\perp}$, respectively. For $f_{v}$, the difference is $-0.78\%$ ($1.18\%$) for Z3H2 (Z2H2), resulting in a maximum value of 254$\%$ of the total allotment. This requires a reduction in the errors by a factor of about 3.2 to get within this 25$\%$ limit.

This could be achieved with a spectroscopic re-observation of some subset of galaxies already observed by \WFIRST\ with a high-resolution ground-based spectrograph that completely resolves H$\alpha$ from the [N\,{\sc ii}] doublet. Only a tiny fraction of \WFIRST\ source could be followed up this way, but it would provide a clean measurement of the redshift error PDF. In this way, we gain knowledge of each re-observed object's specific $z_{\mathrm{true}}$, and can construct a probability distribution $P_{\rm est}(\delta_{z}|z_{\mathrm{obs}})$ based on these re-observed objects. The more objects we are able to re-observe, the more our subsample approaches the full sample, and the more our probability distribution $P_{\rm est}(\delta_{z}|z_{\mathrm{obs}})$ approaches the true distribution $P(\delta_{z}|z_{\mathrm{obs}})$.

In the case of $f_{v}$, we predict how many galaxies would need to be re-observed in order to make this reduction. From Table \ref{tab:finalparams_RSD} it can be seen that (for both Z2H2 and Z3H2):
\begin{equation}
\frac{\partial \mathrm{ln}f_{v}}{\partial \sigma_{\rm FOG}^2} \approx 0.005\,(h^{-1}\,{\rm Mpc})^{-2}.
\end{equation}
This quantifies the dependence of $f_{v}$ on the Finger of God length; if we place limits on $f_{v}$, we need to know $\sigma_{\rm FOG}^{2}$ to some certainty as well, since they are correlated in their fit values. Therefore, to fit within 25$\%$ of the error budget, we need $\sigma_{\rm FOG}^{2}$ to be known to:
\begin{eqnarray}
\sigma_{\rm FOG,err} &=& \left(\frac{\partial \mathrm{ln}f_{v}}{\partial \sigma_{\rm FOG}^2} \right)^{-1} \times f_{v,err} \sqrt{0.25} \nonumber \\
& =& \frac{0.0074\sqrt{0.25}}{0.005} = 0.75 h^{-2}\,\mathrm{Mpc}^{2}.
\end{eqnarray}
Given the variance of the redshift offset $\delta z$, which corresponds to 24.2 $h^{-2}\mathrm{Mpc}^{2}$ for Z2H2 and 21.1 $h^{-2}\mathrm{Mpc}^{2}$ for Z3H2, then we need to know $\sigma_{\rm FOG}^{2}$ to about $0.75/24.2 = 3.1\%$ for Z2H2 and $0.75/21.1 = 3.6\%$ for Z3H2. If assuming Gaussian errors, this would require an observation approximately in the range of $2/0.031^{2} \approx 2100$ galaxies. In practice, the number of re-observations would be larger for a realistic non-Gaussian error distribution, and a few redshift bins would be necessary to track the redshift dependence of the error PDF.

Additional work will be necessary to determine the optimal strategy for measuring the redshift error PDF. Following up thousands of emission-line galaxies with ground-based NIR spectroscopy at the $\sim 10^{-16}$ erg cm$^{-2}$ s$^{-1}$ depth is certainly possible, especially if the targets can be pre-selected from \WFIRST\ to have lines that will not collide with atmospheric OH features. However, since most of the redshift error is statistical error due to photon noise, it may be more efficient to use repeat observations in the \WFIRST\ deep fields to measure the purely statistical scatter, and then high-resolution ground-based NIR spectra to constrain the specific contribution from line blending. These possibilities should be explored in future work.


\section{Discussion}
\label{sec:conclusion}

In this paper, we have examined the effects of the grism resolution proposed for \WFIRST\ on the observed redshifts of galaxies, and the resultant changes in the fitted cosmological parameters $\alpha_{\parallel}$, $\alpha_{\perp}$, and $f_{v}$. We have used the \buzzard\ mock galaxy catalog to probe these parameter differences by simulating the observation of line-blended galaxies and compared them to the true redshift distributions. We then created a ``shuffled'' catalog; this catalog uses the same distribution of $\delta_{z}$ values (see Eq. \ref{eq:deltaz}) as the observed catalog, but with these values randomly shuffled between different galaxies. This results in all correlations between galaxy location and metallicity being erased. By analyzing the differences between the parameters fit to these catalogs, we can gain a sense of the potential parameter errors due to the line blending effect, and to what extent they may require mitigation in order to meet the systematic requirements for \WFIRST.

We found that errors dependent on the large-scale structure, i.e. the difference between the shuffled and observed catalogs, were
$\Delta\alpha_{\parallel} = 0.31 \pm 0.23 \%$, $\Delta\alpha_{\perp} = -0.10 \pm 0.10 \%$, and
$\Delta f_{v} = 0.17 \pm 0.33\%$ ($1.355\le z<1.994$),
$\Delta\alpha_{\parallel} = 0.26 \pm 0.17 \%$, $\Delta\alpha_{\perp} = -0.12 \pm 0.11 \%$, and
$\Delta f_{v} = -0.20 \pm 0.30\%$ ($0.705\le z<1.345$), all quoted at the $1.3\times 10^{-16}$ erg cm$^{-2}$ s$^{-1}$ flux limit.
This uses approximately 18$\%$, 9$\%$, and 7$\%$ of their respective error budgets, in an RSS sense. These errors are small --- in particular, they are smaller than the upper limits presented by other recent analyses \citep{faisst2017} --- and can be made smaller still through the use of mitigation techniques described in Sec.~\ref{sec:mitigation}. Errors that are dependent on the knowledge of the {\em distribution} of galaxy parameters, i.e. the difference between the shuffled and true catalogs, are larger; however these errors are more easily mitigated since the redshift error PDF can be measured by re-observing a small fraction of the sample with high-resolution spectrographs. We estimated that direct mitigation would require re-observation of a few thousand galaxies. The redshift survey C3R2 \citep{2017ApJ...841..111M} will provide samples of high-resolution spectra in \WFIRST\ fields (although the selection criteria are not similar to the \WFIRST\ grism survey). We recommend more work to refine this estimate and define the optimal strategy.

It is important to note that this is only a first study of the effect of [N{\,\sc ii}] + H$\alpha$ line blending on large-scale structure. It examines the effect of one correlation --- the mean dependence of [N{\,\sc ii}]/H$\alpha$ on stellar mass (at fixed redshift), and hence on environment. Future work should investigate a wider range of astrophysical and instrumental sources of bias on and scatter in the observed redshifts, and treat the possible subtle interactions among all these effects. Specific improvements that would be valuable for the next phase of \WFIRST\ studies include:
\begin{list}{$\bullet$}{}
\item {\em Galaxy population}: In this work, we have assumed the environmental dependence is only due to the relationship between galaxy mass and [N{\,\sc ii}]/H$\alpha$ ratio. We expect this to be the dominant effect, both because the additional dependence on SFR (at fixed $M_\star$ and $z$) is observed to be weak, and because there is a strong relation between stellar mass and clustering strength. Future work should explore the robustness of these results under different semi-analytic model assumptions, and different prescriptions for the galaxy SEDs and H$\alpha$ emission line properties. It should also investigate the higher-order moments of the [N{\,\sc ii}]/H$\alpha$ line ratio (e.g., scatter as a function of stellar mass). Realistic correlations of the [N{\,\sc ii}]/H$\alpha$ line ratio with other properties such as the galaxy radius will also be needed to track the interaction of the [N{\,\sc ii}] + H$\alpha$ line blending with other instrument-related biases.
\item {\em Instrument and analysis-induced redshift errors}: In this paper, we have {\em not} included the dependence of the random redshift error on galaxy properties (e.g., size, line flux). Including this is important, but was deferred in this work because it will require significant additional technical steps (particularly in defining the ``shuffled'' catalog). In a grism survey, coma or other forms of PSF asymmetry along the dispersion direction result in redshift biases that depend on galaxy properties (size and emission line $S/N$) as well as field position. While in the real survey these must be addressed with a full survey simulation, a simulation with postage stamps of the emission lines would be a much less computationally demanding intermediate step. This would capture how redshift bias and noise scale with galaxy properties, including dependence on field position and hence the imprint of the tiling strategy (but would not model density-dependent selection effects due to confusion).
\item {\em Survey volume}: In order to investigate higher-order effects that contribute to the error budget, larger survey volume is required. The present study uses a simulated area of $\pi$ sr ($\sim$ 5 times the \WFIRST\ reference survey footprint). In this volume it is hard to measure biases that are small compared to \WFIRST\ statistical errors, even though the differences of true/observed/shuffled catalogs cancel some of the sampling variance. Ideally we would also have enough simulated realizations to build a mock-based covariance matrix (and thus avoid issues related to variable galaxy number density as a function of redshift and non-Gaussianity of the galaxy density fluctuations).
\item {\em BAO reconstruction}: Reconstruction is a non-linear operation on the survey data, and we should search for possible interaction with density-dependent redshift biases.
\end{list}

In summary, we have presented a simulation-based analysis of the effects of [N{\,\sc ii}] and H$\alpha$ line blending on BAO and RSD parameter fitting. We conclude that the errors due to the lowest-order effect (the trend of [N{\,\sc ii}]/H$\alpha$ ratio with stellar mass and hence large-scale environment) are small compared to \WFIRST\ requirements even without mitigation, and with mitigation should not be a concern for these applications of the \WFIRST\ galaxy redshift survey. We have outlined the key improvements needed in future work to study other correlations involving the [N{\,\sc ii}]/H$\alpha$ ratio and their interaction with instrument and analysis-related redshift errors. We have also concluded that the redshift error probability distribution function will need to be measured accurately; while a brute-force approach seems feasible, we recommend further study of the optimal approach.

\section*{Acknowledgements}
We would like to thank Paul Martini for input on galaxy mass--metallicity relations. We appreciate the many useful conversations with Ami Choi, Niall MacCrann, and Heidi Wu. Further thanks to Paulo Montero-Camacho, Benjamin Buckman, Ben Wibking, and Matthew Digman. We thank Ashley J. Ross for his contributions to our correlation function pipeline.
We would also like to thank James Rhoads and Sangeeta Malhotra for providing feedback on our results and taking part in discussions on ideas for future work. We are also grateful for suggestions from an anonymous referee which improved the paper.

DM, XF, and CMH are supported by the Simons Foundation, the US Department of Energy, the Packard Foundation, the NSF, and NASA. RHW and JD received partial support through NASA contract NNG16PJ25C, from NASA ROSES ATP 16-ATP16-0084 grant, and from the U.S. Department of Energy under contract number DE-AC02-76SF00515. This research made use of computational resources at SLAC National Accelerator Laboratory, a U.S.\ Department of Energy Office; the authors thank the support of the SLAC computational team. Many computations in this paper were run on the CCAPP condo of the Ruby Cluster at the Ohio Supercomputer Center \citep{OhioSupercomputerCenter1987}. This research made extensive use of the {\tt arXiv} and NASA's Astrophysics Data System for bibliographic information.


\bibliographystyle{mnras}
\bibliography{sources}
\appendix

\bsp	
\label{lastpage}
\end{document}